\documentclass[10pt,journal,compsoc, twoside]{IEEEtran}

\ifCLASSOPTIONcompsoc
  \usepackage[nocompress]{cite}
\else
  \usepackage{cite}
\fi

\usepackage[utf8]{inputenc}
\usepackage{amsmath}
\usepackage{array}
\usepackage{multirow}
\usepackage{booktabs}
\usepackage{graphicx}
\graphicspath{ {figs/} }
\usepackage{array}
\usepackage{threeparttable}
\usepackage{todonotes}
\usepackage{multirow}
\usepackage{balance}
\usepackage{xcolor}
\usepackage{url}
\usepackage{paralist}
\usepackage{hyperref}
\usepackage{comment}

\begin{document}

\title{A Study about the Knowledge and Use of Requirements Engineering Standards in Industry}

\author{Xavier Franch, Martin Glinz, Daniel Mendez, Norbert Seyff
\IEEEcompsocitemizethanks{\IEEEcompsocthanksitem Xavier Franch is with the Universitat Polit\`ecnica de Catalunya, 08034 Barcelona, Catalonia, Spain. E-mail: franch@essi.upc.edu%
\IEEEcompsocthanksitem Martin Glinz is with the University of Zurich, 8006 Zurich, Switzerland. E-Mail: glinz@ifi.uzh.ch
\IEEEcompsocthanksitem Daniel Mendez is with the Blekinge Institute of Technology, 37179 Karlskrona, Sweden, and also with the fortiss GmbH, 80805 Munich, Germany. E-Mail: daniel.mendez@bth.se
\IEEEcompsocthanksitem Norbert Seyff is with the University of Applied Sciences and Arts Northwestern Switzerland, 5210 Windisch, Switzerland, and also with the University of Zurich, 8006 Zurich, Switzerland. E-Mail: norbert.seyff@fhnw.ch}
\thanks{Manuscript received 25 Jan. 2021; revised 14 May 2021; accepted 28 May 2021.\\
Recommended for acceptance by Z. Jin.\\
Digital Object Identifier no. 10.1109/TSE.2021.3087792}}

\markboth{IEEE TRANSACTIONS ON SOFTWARE ENGINEERING}%
{Shell \MakeLowercase{\textit{et al.}}: Bare Demo of IEEEtran.cls for Computer Society Journals}
%

\IEEEtitleabstractindextext{%
\begin{abstract}
\emph {Context.} The use of standards is considered a vital part of any engineering discipline. So one could expect that standards play an important role in Requirements Engineering (RE) as well. However, little is known about the actual knowledge and use of RE-related standards in industry.
\emph {Objective.} In this article, we investigate to which extent standards and related artifacts such as templates or guidelines are known and used by RE practitioners.
\emph {Method.} To this end, we have conducted a questionnaire-based online survey. We could analyze the replies from 90 RE practitioners using a combination of closed and open-text questions.
\emph {Results.} Our results indicate that the knowledge and use of standards and related artifacts in RE is less widespread than one might expect from an engineering perspective. For example, about 47\% of the respondents working as requirements engineers or business analysts do not know the core standard in RE, ISO/IEC/IEEE 29148. 
Participants in our study mostly use standards by personal decision rather than being imposed by their respective company, customer, or regulator.
Beyond insufficient knowledge, we also found cultural and organizational factors impeding the widespread adoption of standards in RE.
\emph {Conclusions.} Overall, our results provide empirically informed insights into the actual use of standards and related artifacts in RE practice and – indirectly – about the value that the current standards create for RE practitioners.
\end{abstract}

\begin{IEEEkeywords}
Standard, requirements engineering, guideline, template, survey
\end{IEEEkeywords}}

\maketitle

\section{Introduction}\label{sec:introduction}
Standards play a vital role in engineering, especially when many manufacturers, suppliers, and providers need to collaborate and interact for developing products and providing services in given domains.
The use of standards allows ``to benefit from the wisdom and experience of others, rather than reinvent the wheel each time'' \cite{Osif2014}. Harding and McPherson~\cite{HaMcP2010} found that ``Managers recognize standardization as essential to daily operations, regardless of the company’s economic sector''. Consequently, knowing standards and learning how to work with them can be considered relevant in engineering education~\cite{ABET2017}\cite{Phillips2019}.

Being an engineering discipline, one could expect that standards play an important role in Requirements Engineering (RE) as well. However, we have personally observed over the years that there are people in industry who do not use any standards or who only use them scarcely in their RE work. Moreover, practitioners (as well as researchers) may lose track of the evolution of standards in the field. For instance, we still find today publications citing the IEEE 830-1998 standard~\cite{IEEE830_1998} although it has been retired and superseded by the ISO/IEC/IEEE 29148 standard~\cite{ISO29148}. Hence, our observations indicate that RE practitioners might not widely adopt or even know RE-related standards.

This motivated us to investigate the role of standards in contemporary RE. This article reports on a survey study about the degree to which RE-related standards are actually known and used in industry---and why they are or are not used. To the best of our knowledge, there is no systematic study of this topic so far.

The results of our study contribute insights into the actual use of RE standards in industry. We found that the knowledge and use of standards in RE as reported in the study is less widespread than one might expect. Furthermore, we identified which standards are most known and used by practitioners, and whether and how they use other types of artifacts with the same purpose as standards.

All data analyzed and reported in this article is available as an open replication package under \url{http://doi.org/10.5281/zenodo.4755756} including codebook and instrument, raw data and filtered data, and the results of our data analysis with respect to the research questions and the demographics.

\markboth{IEEE TRANSACTIONS ON SOFTWARE ENGINEERING}{FRANCH ET AL.: STUDY ABOUT THE KNOWLEDGE AND USE OF REQUIREMENTS ENGINEERING STANDARDS IN INDUSTRY}

\section{Background and Related Work}\label{sec:background}
In the context of engineering, the term \emph{standard} refers to
``something established by authority, custom, or general consent as a model or example or (...)  as a rule for the measure of quantity, weight, extent, value, or quality'' (Merriam-Webster online dictionary). This definition also applies to software engineering in general and RE in particular, where standards may prescribe methods, templates, quality metrics and definitions of terms.

Using standards in RE has many advantages. Adhering to standards makes work products less dependent on the individual capabilities and working styles of the people producing these work products. It also makes work products easier to read and comprehend, particularly for people not involved in creating them. Frequently, standards provide procedures or predefined structures. Adhering to them helps avoid mistakes and improves the quality of the work products produced.

Most RE-related standards are created and published by recognized standardization organizations such as ISO, IEEE, or OMG. We call these \emph{de-jure} standards. Besides, there are also  \emph{de-facto} standards which are created by companies, authors of influential books, or organizations of professionals and widely adopted by the community. 
We further distinguish between \emph{normative standards} about practices, documentation, and terminology on the one hand, and \emph{standardized artifacts}, e.g., languages, templates or guidelines (such as UML or VOLERE) on the other hand.
Schneider and Berenbach~\cite{ScBe13} conducted a literature survey on \emph{de-jure} and normative RE-related standards including the eight most adopted standards at that moment (2013), together with their relationships and some practical observations and advices on their use. We have not found any follow-up to this literature survey, neither updating the collection of \emph{de-jure} and normative standards, nor including \emph{de-facto} standards or standardized artifacts.

 The use of RE-related standards in practice has not been studied so far. Khan et al. report that there are no systematic mapping studies conducted on software engineering standards in general (thus, RE in particular)~\cite{Khan2019}. Still, some studies do mention occasionally standards as part of a more general goal. The NaPiRE study~\cite{Mendez2017}\cite{Wagner2019} includes an investigation of whether or not companies have defined RE process standards. While most of the analyzed projects did, only 25\% of them used standards predefined by a regulation. The RE-Pract study~\cite{FranchTSE} reports some observations from practitioners about the perceived importance of standards, e.g., ``Aligning requirements to regulatory standards is extremely important in many industry", ``I recomend to use the IEEE's standard, that was a better guidance to achive my objective"  [{\it sic}]. However, neither NaPiRE nor RE-Pract went any further in that direction.
 
 Our study aims at closing this gap. We are interested in the industrial adoption of all types of standards mentioned above.  This interest is reflected in the research questions, and in the subsequent survey questions presented next.

\section{Study Design}\label{sec:study}

\subsection{Goal and Research Questions}
The overall goal of our study is \emph{to understand to which extent RE practitioners know standards, how often they use them, and what the underlying reasons are}. To this end, we conducted a questionnaire-based online survey. 
The scope of our study encompasses (a)~specific RE standards, for example, ISO/IEC/IEEE 29148~\cite{ISO29148}, 
and (b)~standards from other fields such as quality or systems and software engineering that discuss aspects of RE processes, practices, and artifacts. We use the term \emph{RE-related standards} for denoting the standards in the scope of our study and include both de-jure standards and de-facto ones. Furthermore, we also target \emph{standardized artifacts} such as UML diagrams or use case templates, since we are aware that many practitioners use them with a similar purpose as standards.

We derive four main research questions (RQs), listed in Table~\ref{RQ-table}. RQ1, RQ2, and RQ3 focus on the knowledge and use of RE-related standards and, thus, they are useful for standardization organizations as well as for practitioners who might want to better understand or adopt standards. RQ4 investigates the use of standardized artifacts, in particular, guidelines or templates. With this question, we anticipate the case where some practitioners might not be using any normative standards, but adhere to well-established practices or processes that may have been issued, e.g., by their company.

\begin{table}[!t]
\renewcommand{\arraystretch}{1.3}
\caption{Research questions and sub-questions.}
\label{RQ-table}
\centering
\begin{tabular}{lp{7cm}}
\textbf{RQ1} & \textbf{To what extent are RE-related standards known by practitioners?
} \\
RQ1.1 & What RE-related standards are known by practitioners? \\
RQ1.2 & How did practitioners know about these standards? \\
\textbf{RQ2} & \textbf{To what extent are RE-related standards used by practitioners?} \\
RQ2.1 & How often do practitioners use each of these standards? \\
RQ2.2 & For which purpose do practitioners use each of these standards? \\
RQ2.3 & Why do practitioners use these standards? \\
RQ2.4 & Why do practitioners not use the standards that they know? \\
\textbf{RQ3} & \textbf{Under which conditions are RE-related standards used by practitioners?} \\
RQ3.1 & What factors affect this usage? \\
RQ3.2 & What are the major impediments in effectively using RE-related standards? \\
RQ3.3 & What are the perceived advantages coming from the use of RE-related standards? \\
\textbf{RQ4} & \textbf{Do practitioners use further guidelines or documentation templates in their RE process?} \\
RQ4.1 & What type of guidelines or templates do they use? \\
RQ4.2 & How often are these guidelines or templates used? \\
RQ4.3 & For which purpose do practitioners use each of these guidelines or templates? \\
RQ4.4 & How are these guidelines or templates  used in combination with RE-related standards? 
\end{tabular}
\end{table}

\subsection{Selecting Standards for the Survey}
\label{sec_select_std}

We decided to provide a list of candidate standards where respondents can mark the ones they know. This approach has the disadvantage of potentially biasing the respondents’ answers by our pre-defined list of standards. Yet, we consider this approach to be much faster and more convenient for respondents than forcing them to write down the names of the standards which could make respondents forget about some of the standards they know (rendering comparison of responses invalid). Another advantage is that when respondents do not check a standard in the presented list of standards, we can conclude that they actually do not know it.

To mitigate bias, we decided to present respondents with both a pre-defined list of RE-related standards and a list of empty text fields, where respondents could enter additional standards that they know and consider relevant, e.g., domain-specific standards that they deem relevant for their RE work.

We chose standards according to scope (including RE-related standards from the fields of \emph{quality} and \emph{systems and software engineering} but not considering \emph{domain-specific} standards) and type (both {\it de-facto} and {\it de-jure}). The first and obvious candidate is \emph{ISO/IEC/IEEE  29148} \cite{ISO29148}, which is primarily about RE processes and documentation. We used the 2011 version, as the revised version (2018) was not yet available when we designed the survey. We also included \emph{IEEE 830-1998}~\cite{IEEE830_1998} on software requirements specifications. Although IEEE 830-1998 has been retired and superseded by ISO/IEC/IEEE 29148, we had sufficient evidence that it is still in use.

Among the ISO standards, we found \emph{ISO/IEC TR 24766}~\cite{ISO24766} on RE tool capabilities, and \emph{ISO/IEC 26551} \cite{ISO26551} on product line requirements engineering. As both of them deal with a specific subtopic in RE, we selected the first one as a representative for this category. 

When searching for non-official industry standards, we found the \emph{IREB Glossary of RE Terminology}~\cite{IREB_Glossary}. Given the fact that using this glossary is mandatory in the IREB CPRE certification and more than 50,000 people world-wide hold an IREB CPRE Foundation Level certificate, we consider this glossary a de-facto standard for RE terminology. 

Next, we searched the field of \emph{quality} standards for requirements-related ones. We chose \emph{ISO/IEC 25010}~\cite{ISO25010} on system and software quality models and  \emph{ISO/IEC 25030}~\cite{ISO25030} on quality requirements as representatives. Furthermore, we included \emph{ISO/IEC 9126}~\cite{ISO9126}. Although this family of standards has been superseded by the ISO/IEC 250xx family, it is still used and referenced.

Finally, we searched for RE-related standards in the field of \emph{systems and software engineering}. We included \emph{ISO/IEC/IEEE 12207}~\cite{ISO12207} on software life cycle processes (2008 and 2017 versions) and \emph{ISO/IEC/IEEE 15288}~\cite{ISO15288} on system life cycle processes: both of them are closely related to ISO/IEC/IEEE 29148. Further, we decided to include another terminology standard, \emph{ISO/IEC/IEEE 24765} ~\cite{ISO24765} on systems and software engineering vocabulary (2010 and 2017 versions).

Table \ref{tab:SelectedStandards} lists the selected standards. More details about standards in RE and how they relate to each other can be found in Schneider and Berenbach's paper~\cite{ScBe13}.

\begin{table*}[htb]
\caption{Standards selected for the questionnaire.}
    \label{tab:SelectedStandards}
    \centering
\renewcommand{\arraystretch}{1.3}
\begin{tabular} { l l l l }
\toprule
ISO/IEC/IEEE 29148 & current & RE specific & RE processes and documentation \\
IEEE 830-1998 & superseded & RE specific & RE processes and documentation \\
ISO/IEC TR 24766 & current & RE specific & RE tool capabilities \\
IREB Glossary & current & RE specific & RE vocabulary \\
ISO/IEC 25010 & current & Quality & Systems and Software quality models \\
ISO/IEC 25030 & current & Quality & Quality requirements \\
ISO/IEC 9126 & superseded & Quality & Systems and Software quality models \\
ISO/IEC/IEEE 12207 & current & Systems and Software Engineering & Software life cycle models \\
ISO/IEC/IEEE 15288 & current & Systems and Software Engineering & System life cycle models \\
ISO/IEC/IEEE 24765 & current & Systems and Software Engineering & Sys\&Sw engineering vocabulary \\
\bottomrule
\end{tabular}
\end{table*}

\subsection{Target Population}
\label{sec_select_participants}
 
As a target population, we chose practitioners who act or who have acted in the role of a requirements engineer in at least one industry project, and also practitioners from neighbouring fields such as quality management or testing. That is to say, we used the term \emph{requirements engineer} in its broadest possible sense~\cite{FranchCACM}, including technical and non-technical roles dealing with the specification and management of requirements. We relied on convenience sampling as described in Sect.~\ref{sec:datacollection}.

\subsection{Instrument}\label{subsec:instrument}

Our survey questionnaire starts with questions for gathering demographic information with the purpose of analyzing it later to identify factors which may have an influence on the choices made by our respondents. In particular, we implemented the following demographic factors: country of work; highest education degree; job functions performed in the last six years; years of experience in RE or related fields; RE-related certificates they hold (e.g., from IREB\footnote{\url{https://www.ireb.org/}} or ISTQB\footnote{\url{https://www.istqb.org/}}); average number of people involved in their projects; organizational role played by the team in the company; main industrial sector;
class of systems delivered. 

The rest of the survey included one question per sub-RQ. First, respondents provided their knowledge of the selected standards (as described in Sect.~\ref{sec_select_std}). Next, respondents were able to add standards they know in addition to those given as pre-defined options for selection. Those added standards were included in follow-up questions to answer RQ2. For RQ2, the questionnaire was dynamic with respect to previous answers. For example, the survey presented only those standards to the respondents that they had declared to know in RQ1.1.

\subsection{Data Collection}
\label{sec:datacollection}
To enable world-wide participation, we decided to use a web-based survey tool. To this end, we implemented the survey using the Questback survey tool. 
After the initial implementation of the questionnaire, three RE practitioners were invited to fill it out as part of a piloting phase to ensure its correct understanding and to find possible defects regarding content and presentation. Based on these pilots, we implemented some minor corrections (e.g., by removing duplicates in the answer options), added some information (e.g., how to interpret the questions), and also made some minor text changes.

Data collection took place over six weeks starting by February 2019. In order to increase our target audience, we used various social media channels and mailing lists, e.g. the IREB mailing list. Moreover, we asked in our personal networks of practitioners to spread the word, which means that we cannot communicate a response rate. To compensate for the threats emerging from this convenience sampling strategy, we included a broad set of demographic questions (see Sect.\ref{subsec:instrument}).

\subsection{Data Analysis}\label{sec:DataAn}

For the analysis of RQ1 and RQ2, we exported the survey results to spreadsheets and calculated descriptive statistics. In all the sub-questions, we provide first an overall view of the results in shares of responses relative to the total number of responses. We use accompanying figures to illustrate these global results. Moreover, in RQ2.2, RQ2.3 and RQ2.4, we provide additional insights related to individual standards as shares of respondents that use (RQ2.2, RQ2.3) or do not use (RQ2.4) such standards. Finally, we remark that we have varying numbers of responses as reported in Table~\ref{tab:SQ-Responses}. This is due to the fact that we used dynamic questions for RQ2, as explained in Section~\ref{subsec:instrument} above.

RQ3 required the analysis of free text answers. We applied content analysis according to the following steps:
\begin{compactenum}
\item \emph{Filtering.} Responses were filtered by excluding: a) empty fields (including responses as ``---" or ``n.a."); b) text not answering the questions  (e.g., ``Regularly" as answer to the question ``What are the factors driving the use of the standards?"); c) vagueness (e.g., ``To do better job"); d) lack of expertise (e.g., ``I'm not close enough to the standards used to answer this question confidently"). A few long responses were kept as valid but some parts fall into the cases b)-d), in which case we removed the affected part, but kept the rest.
\item \emph{Term extraction.} The remaining responses were split into semantically relevant terms. In this step, only equal or very similar terms (e.g., ``Lack of awareness" and ``Limited awareness") were aligned. Non-essential, very concrete information was also disregarded (e.g., ``Keep CMMI certification" was converted into ``Company certification").
\item \emph{Topic identification.} The extracted terms were compared to each other looking for topics that reflect semantic similarities. For instance, the topic ``Lack of experts" includes the terms ``Lack of training" and ``Lack of consultancy". Some methodological guidelines follow:
\begin{compactitem}
\item We analyzed the context of every sentence. For instance, the impediment term ``Ignorant managers" could be classified only as ``Company capacity" (namely, managers do not have the skills to use standards in RE activities) but a careful reading of the context shows that it also refers to an attitude problem, so it needs to be classified also into the topic ``Employees' attitude".
\item We avoided over-interpretation; instead, we preferred to reflect the respondents' opinions. For instance, two benefits mentioned by respondents were ``Acquisition of a shared language" and ``Improved communication". While the former is a means for the latter, we preferred to keep them separated given the different emphasis.
\end{compactitem}
\item \emph{Category identification.} The identified topics were grouped into categories to facilitate further analysis. For instance, the category ``Project factors" includes the topics ``Time", ``Cost", ``Effort" and ``Risk".
\end{compactenum}
The coding was performed by one of the authors and reviewed by another author. The coder and the reviewer discussed and resolved the findings from the review.

RQ4 followed the same four steps as RQ3. However, given the higher diversity of responses that soon emerged, we followed an iterative process:
\begin{itemize}
    \item In a first iteration, two authors analyzed the baseline data, with focus on filtering and term extraction. Then, they held a first dedicated meeting to discuss their proposals and understand each other's rationale and brainstorm about topics and categories.
    \item Building on this shared understanding, one of these two authors conducted a second iteration with a complete proposal on filtering and term extraction and a draft of topics and categories, which was consolidated in a meeting among the two authors.
        \item In a subsequent plenary meeting, all authors discussed the coding proposal and identified some issues with the codes and the categories.
    \item Last, a third author made a last iteration to adjust the proposal, which was again discussed by all authors, leading to the final coding.
\end{itemize}

\section{Results}\label{sec:results}
In this section, we report the results related to the four research questions, preceded by demographics of the study. Throughout the section, we will use the notation ``[N;P\%]" to inform about statements provided by \emph{N} respondents amounting to \emph{P\%} of the total number of respondents answering that particular question. Besides, unless otherwise stated, graphics will show absolute numbers in the Y-axis and percentages over the corresponding bars. Please note that while all respondents provide answers to the questions under RQ1, the number of respondents to RQ2 varies as these answers are given only to those standards the respondents know and use (or don't use), and the variations in RQ3 and RQ4 are even greater since not all respondents provided answers to the open text questions. Therefore, we will specify the population size explicitly as to make the percentages accurate.

\subsection{Demographics}\label{subsec:Demo}
We received in total 99 answers from which we discarded 9 either because they were empty or because they contained dubious information. Demographic characteristics are as follows:

\emph{Country of work}.
Respondents are spread over all continents except Africa with Europe having the biggest share [48;53.3\%]. China and Germany are the dominant countries with [19;21.1\%] respondents for each of them. 

\begin{figure}[!t]
\centering
\includegraphics[width=1\columnwidth]{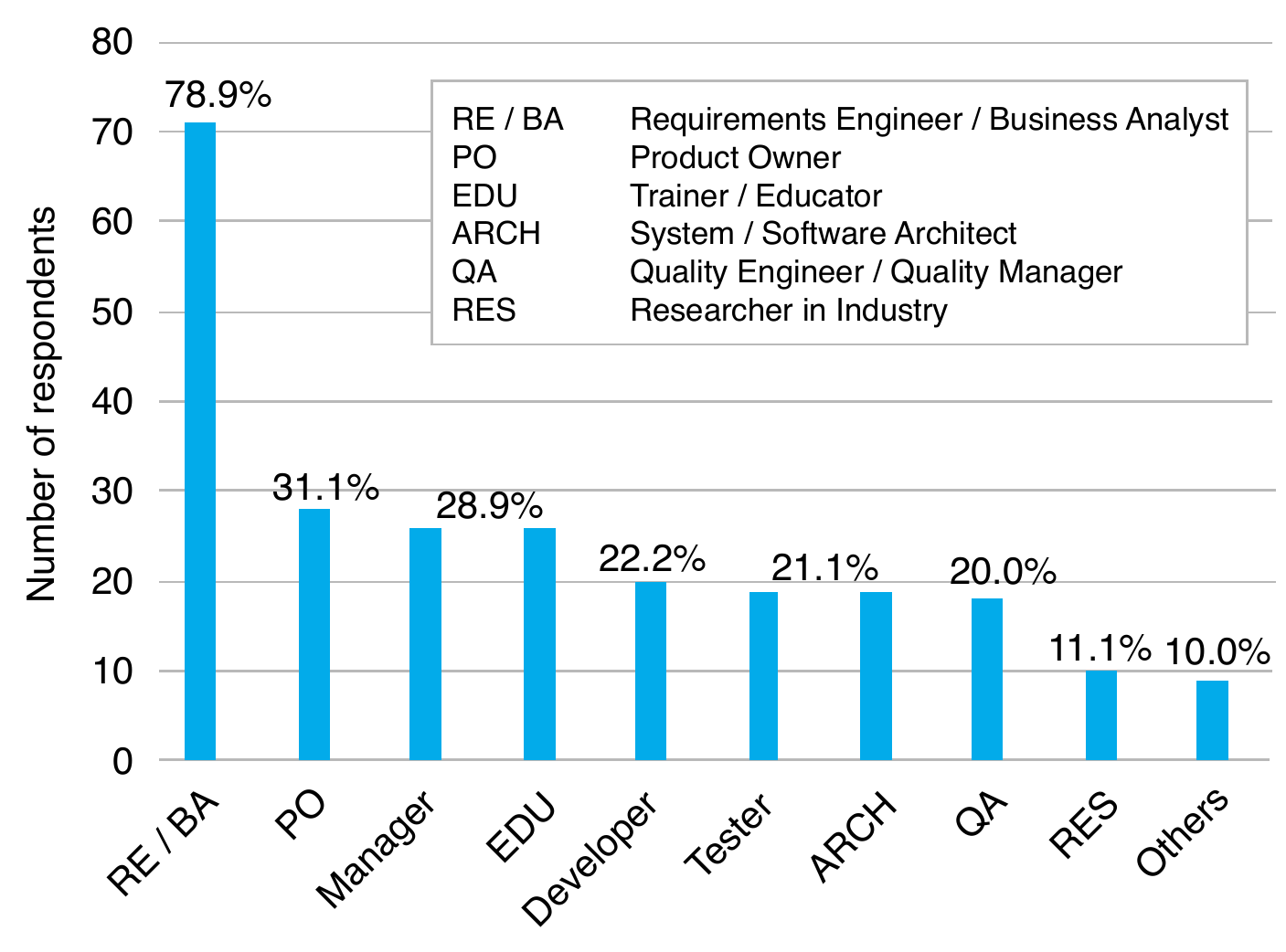}
\caption{Demographics: Job functions of respondents in the last six years.}
\label{demographics-position}
\end{figure}

\emph{Position}. Most of the respondents declare more than one job function in the last six years [68;75.6\%]. Among them, the most popular is by far the job of requirements engineer/business analyst reaching almost 80 percent of respondents, with the rest 
distributed rather uniformly except for researcher in industry (see Figure \ref{demographics-position}).
A few respondents report a handful of other positions, as consultant or certification specialist.

\emph{Certificates}. Answers to this open-text question were very diverse
\footnote{Some responses were a bit ambiguous and we needed to take some decisions to resolve these ambiguities.}.

Many  respondents hold a professional training certificate  [52;57.8\%],  some have even  more  than  one [22;24.4\%]. A slight majority of respondents hold an RE-related certificate [46;51.1\%], with IREB certificates being by far the most popular ones ([43;47.8\%] when considering all types of certificates). The runner-up is ISTQB certification [11;12.2\%].
A majority of respondents certified by these two associations are at the foundation level (IREB-FL and ISTQB-FL) [44;48.9\%], while the number of respondents with advanced or specialist level certificates is lower [13;14.4\%]. 
A few respondents have some agile-related certificate [7;7.8\%]. Still a good number of respondents [19; 21.1\%] have certificates not in the former categories, with CBAP, PMI-PBA, CCBA and TOGAF mentioned by more than one respondent.

\emph{Years of experience}. The majority of respondents have ten years or more of experience [39;43.3\%] or between four and nine years [33;36.7\%], which can be considered a good indicator for the quality of the information gathered in the questionnaire. 

\emph{Education.} The majority of respondents hold an MSc degree  [39;43.3\%]\footnote{Including six people with a German Dipl.-Ing. degree.} or a BSc degree [29;32.2\%]. [9;10.0\%] have a PhD. [7;7.8\%] have a diploma or engineering degree whose equivalence to BSc or MSc cannot be established. [6;6.7\%] hold other or unspecified degrees.


\emph{Project size}. 
The majority of respondents works in small teams with up to ten members [44;48.9\%], while a similar amount works in teams with more than ten but less than hundred members [40;44.4\%]. Only a few respondents work in very large teams with one hundred or more members (up to eight hundred) [6; 6.7\%].

\emph{Main industrial sector}. No single sector dominates among the respondents. \emph{Finance} is the most mentioned sector [14;15.6\%], followed by \emph{Automotive} [10;11.1\%], \emph{Healthcare} [10;11.1\%], \emph{Manufacturing} [8;8.9\%], \emph{E-commerce} [7;7.8\%], and \emph{Energy} [6;6.7\%]. The remaining 19 mentioned sectors do not exceed four responses each. A few respondents [4;4.4\%] report that they cannot identify a single sector.

\emph{System class}. Nearly half of the respondents [42;46.7\%] work mainly on business information systems. Twenty respondents work on software-intensive embedded systems   [20;22.2\%] and also twenty on hybrid systems. A few respondents work on other types of systems [8;8.9\%], such as mobile apps or research software development.

\subsection{RQ1. To what extent are RE-related standards known by practitioners?}
\label{sec_RQ1}
\subsubsection{RQ1.1. What RE-related standards are known by practitioners?}
\label{RQ1_1}

\begin{figure}[!b]
\centering
\includegraphics[width=\columnwidth]{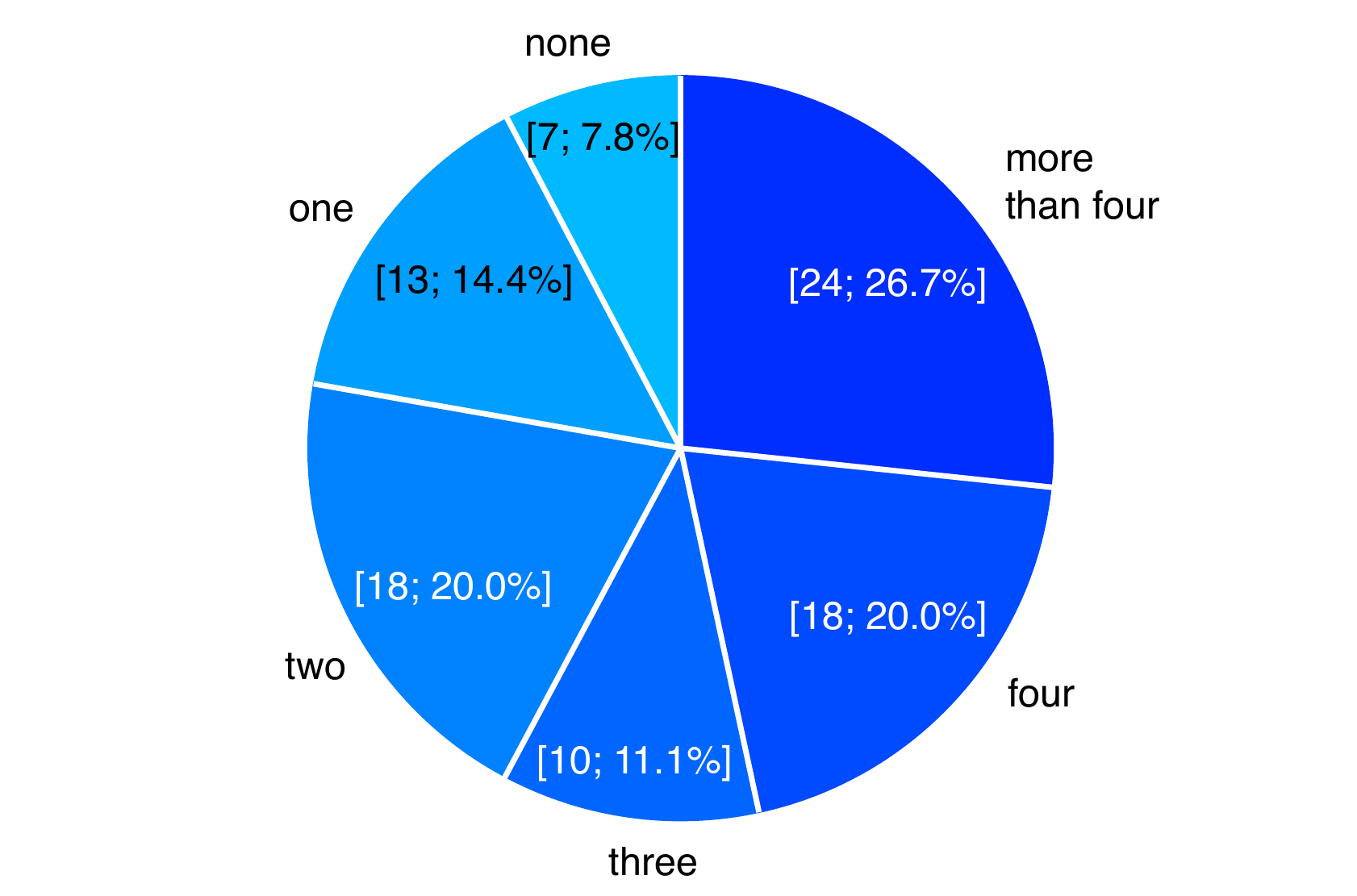}

\caption{RQ1.1: Number of RE-related standards known.}
\label{fig:standards-numbers}
\end{figure}

Fig.~\ref{fig:standards-numbers} shows how many RE-related standards the respondents know. These numbers include the ten standards we had enumerated in the survey, as well as the further standards that the respondents specified in the free form entry field. Numbers are diverse. While about 27\% of respondents know more than four standards, a larger percentage (over 42\%) do not know more than two, with almost 8\% saying that they do not know any.

\begin{figure}[!t]
\centering
\includegraphics[width=\columnwidth]{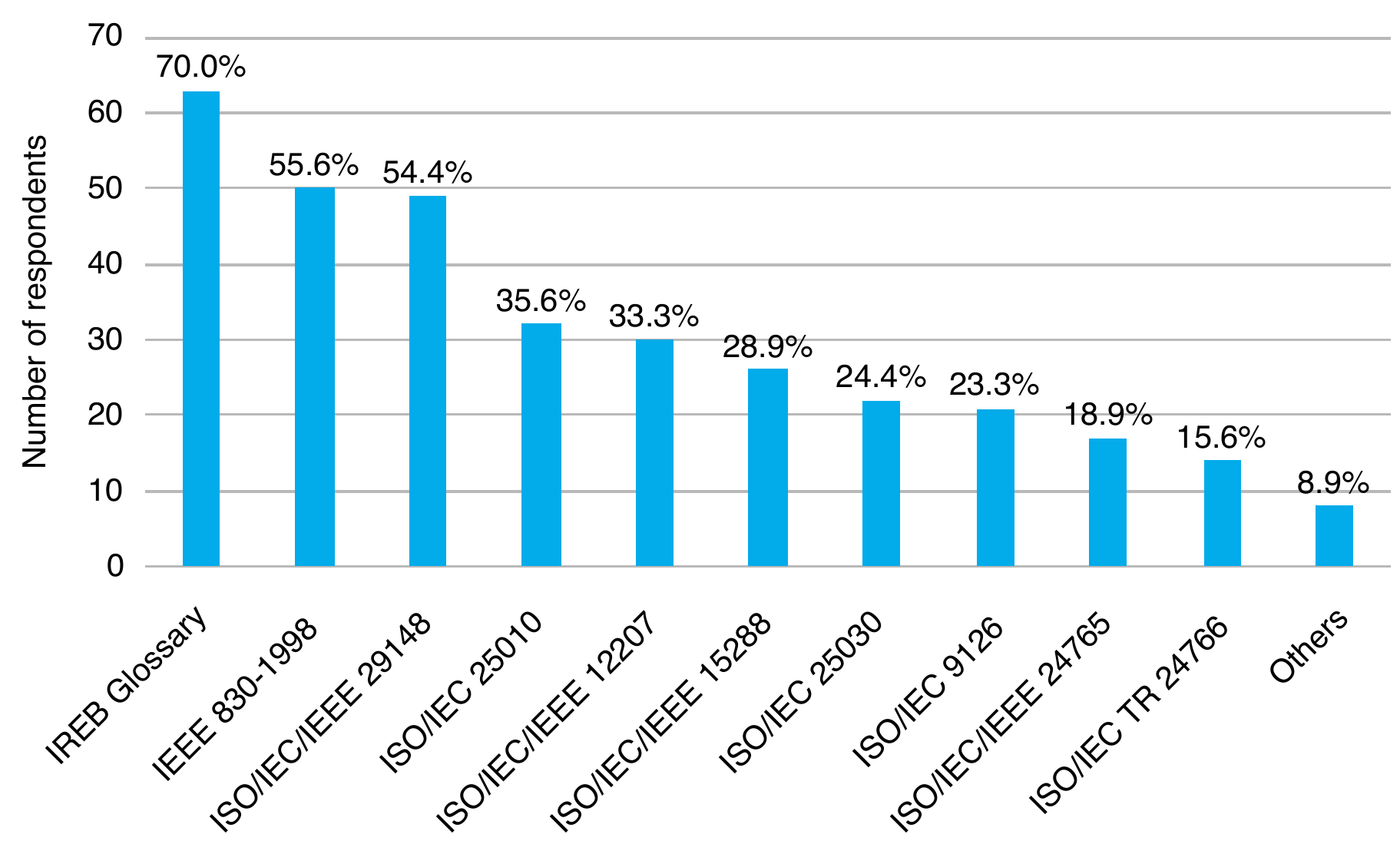}
\caption{RQ1.1: Knowledge of RE-related standards (percentages given over the set of 90 respondents).}
\label{fig:standards-ranking}
\end{figure}

Fig.~\ref{fig:standards-ranking} shows the knowledge of the ten RE-related standards listed in Table \ref{tab:SelectedStandards} by the 90 respondents. The most widely known standard is a {\it de facto} standard, the IREB Glossary of Requirements Engineering Terminology, known by 70\% of respondents. 
This number aligns with the large share of respondents who are certified by IREB (see Sect.~\ref{subsec:Demo}). 
The other two standards that excel in the responses are the two core RE {\it de-jure} standards, ISO/IEC/IEEE 29148~\cite{ISO29148} and its predecessor, IEEE 830-1998~\cite{IEEE830_1998}, which are known by about 55\% percent of the respondents. Less than 36\% of the respondents know the remaining standards.

In response to our open question asking for additional RE-related standards, roughly one third of the respondents [29;32.2\%] suggested up to 38 terms or concepts. However, after analyzing the suggestions, only seven of them can be considered as RE-related standards: IEC 62304 on Medical Device Software~\cite{IEC62304}, ISO 16404 on requirements management for space systems~\cite{ISO16404}, the aircraft standards ARP4754A~\cite{ARP4754A} and DO-178~\cite{DO-178C}, ISO/IEC 29110 on profiles and guidelines for Very Small Entities~\cite{ISO29110}, IEEE 1233-1998, a guide for developing SRS~\cite{IEEE1233}, and ISO/IEC 26551 on product line RE~\cite{ISO26551}.

\subsubsection{RQ1.2. How did practitioners know about these standards?}
\begin{figure}[!t]
\centering
\includegraphics[width=\columnwidth]{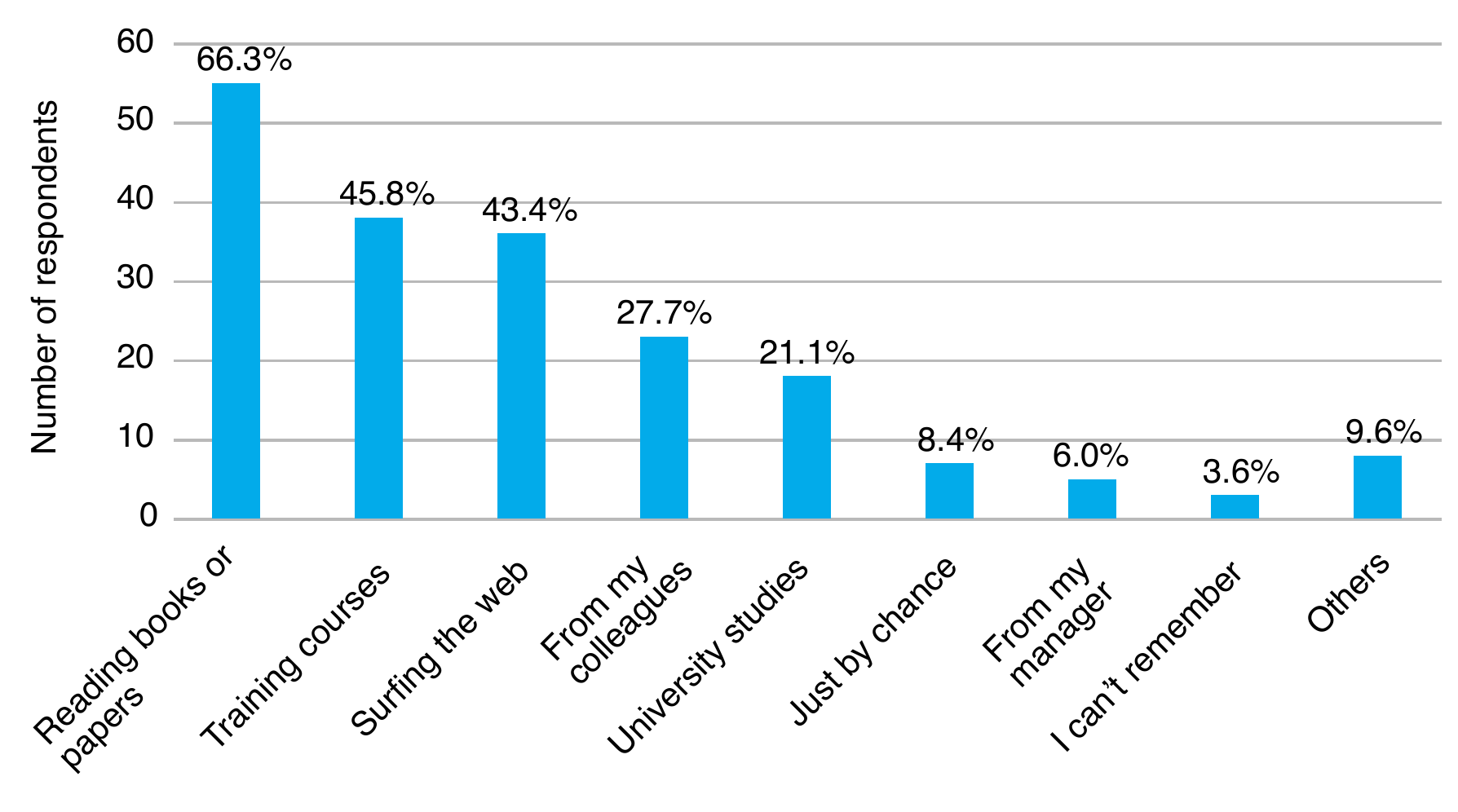}
\caption{RQ1.2: Sources for the knowledge of RE-related standards  (percentages given over the set of 83 respondents who know at least one standard).}
\label{fig:standards-sources}
\end{figure}

 All the 83 respondents who know some standard answered this question\footnote{Therefore, percentages in RQ1.2 are computed over the 83 respondents who know some standard.}. Almost three quarters of them [60;72.3\%] know standards by more than one source, and even a few [10;12.0\%] from four or more sources. Fig. \ref{fig:standards-sources} summarizes the responses for the sources presented in the questionnaire. Books and papers are 
the dominant source mentioned by two thirds of respondents, while the other two remarkable sources are training courses and the web. A few respondents mentioned sources not in the list, e.g., ``From our customers".

\subsection{RQ2: To what extent are RE-related standards used by practitioners?}
As mentioned in Sect.~\ref{subsec:instrument}, the questionnaire was dynamic in this part, showing the respondents only those standards they selected in previous questions: 
\emph{(i)} RQ2.1 asked only about those standards that the respondents had declared to know under RQ1.1; \emph{(ii)} RQ2.2 and RQ2.3 asked only about those standards that the respondents had declared to use in at least some projects under RQ2.1; \emph{(iii)} RQ2.4 asked only about those standards that the respondents had declared to know but \emph{not} to use in any project under RQ2.1. Table~\ref{tab:SQ-Responses} shows the resulting numbers\footnote{A few respondents did not provide information, therefore numbers do not completely match with RQ1's results.}. We discarded for analysis the standards mentioned in the  free-text responses, because they were mentioned only by two respondents at most.

\begin{table}[h]
\caption{RQ2. Responses to survey questions}
    \label{tab:SQ-Responses}
    \centering
\renewcommand{\arraystretch}{1.3}
\begin{tabular} {c | c c c c c c c c c c }
\rotatebox[origin=c]{90}{\emph{~Research Question~}} & \rotatebox[origin=c]{90}{\emph{~IEEE 830-1998~}} & \rotatebox[origin=c]{90}{\emph{~ISO/IEC/IEEE 29148~}} & \rotatebox[origin=c]{90}{\emph{~IREB Glossary~}} & \rotatebox[origin=c]{90}{\emph{~ISO/IEC TR 24766~}}  & \rotatebox[origin=c]{90}{\emph{~ISO/IEC 25010~}} & \rotatebox[origin=c]{90}{\emph{~ISO/IEC 25030~}} & \rotatebox[origin=c]{90}{\emph{~ISO/IEC 9126~}} & \rotatebox[origin=c]{90}{\emph{~ISO/IEC/IEEE 24765~}} & \rotatebox[origin=c]{90}{\emph{~ISO/IEC/IEEE 12207~}} & \rotatebox[origin=c]{90}{\emph{~ISO/IEC/IEEE 15288~}}\\
\hline
2.1      & 50 & 49 & 63 & 14 & 32 & 22 & 21 & 17 & 30 & 26 \\
2.2\&2.3 & 39 & 34 & 46 &  9 & 22 & 14 & 12 & 14 & 24 & 21 \\
2.4      &  8 & 12 & 14 &  4 &  9 &  6 &  8 &  2 &  2 &  2 \\
\end{tabular}
\end{table}

\begin{figure}[!htb]
\centering
\includegraphics[width=\columnwidth]{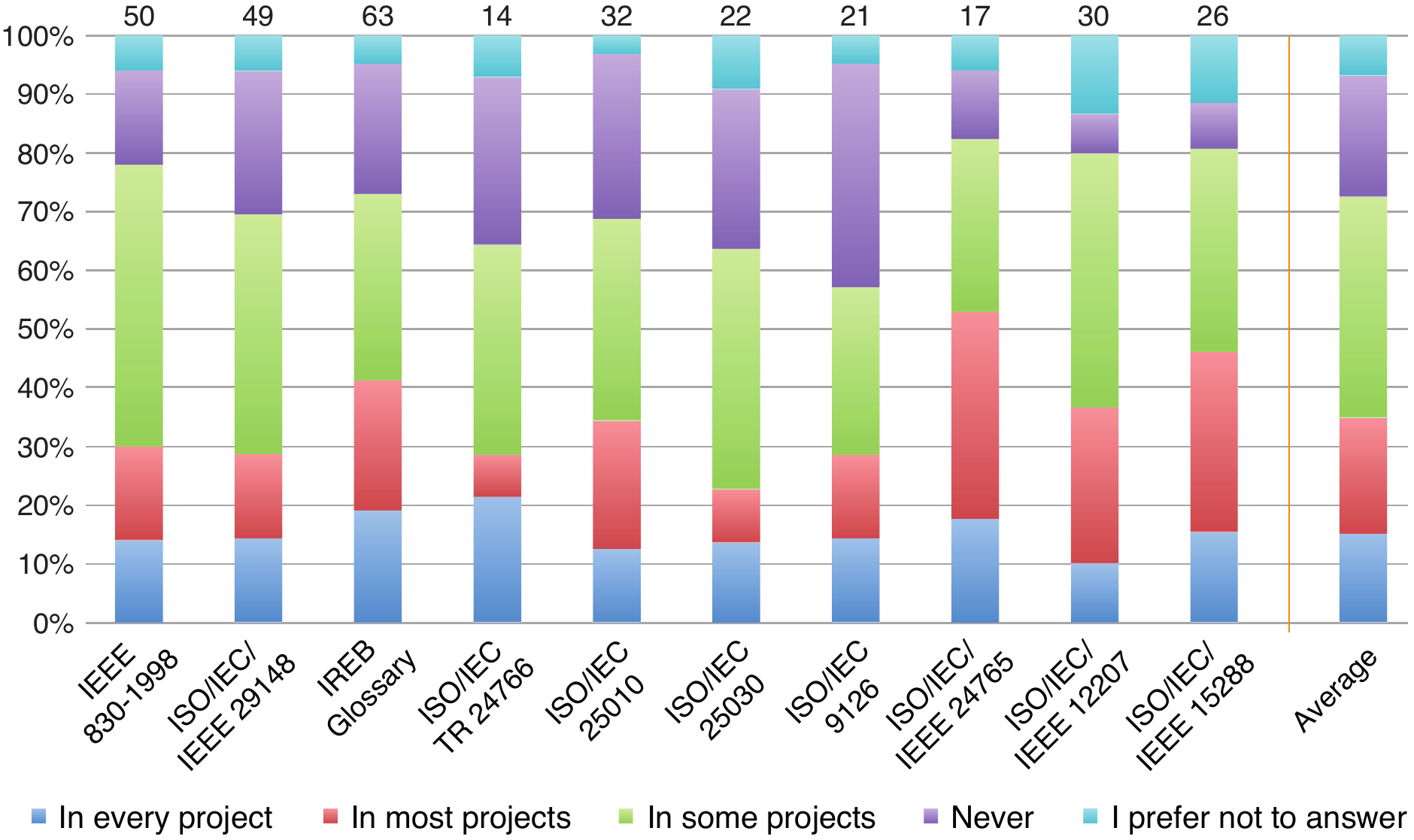}
\caption{RQ2.1: Frequency of using RE-related standards (number of respondents knowing each standard as per RQ1.1 shown above the bars).}
\label{standards-frequency}
\end{figure}

\subsubsection{RQ2.1: Frequency of using standards}
Fig.~\ref{standards-frequency} summarizes the frequency of using the standards. For each standard, we show the percentages of use relative to the number of respondents who had stated to know the standard. From the great amount of results we could compare, we have selected two particular observations to report.

\emph{1.~Which standards are used intensively (i.e., in every or in most projects)?} Although all known standards are used at least in some projects, the percentage of respondents who use them in every project is rather low (only 15.1\% of responses) and no single standard excels with respect to this continuous usage (results range from 10.0\% in the case of ISO/IEC/IEEE 12207 to 21.4\% for ISO/IEC TR 24766). If we add responses that report usage in most projects, we can observe that the numbers grow to 34.9\% on average
with ISO/IEC/IEEE 24765 [9;52.9\%] and ISO/IEC/IEEE 15288 [12;46.2\%] being the two most frequently used standards in terms of percentages, while the others gradually decrease in percentage down to ISO/IEC 25030 [5;22.7\%]. If we focus on absolute numbers, the IREB Glossary is the most intensively used standard with 26 respondents, which represents 41.2\% of the respondents who know it.

\emph{2.~Which standards are never used?} 
The ISO/IEC 9126 standard is the least used one in our study: it is never used by eight 
of the 21 respondents who know it [8;38.1\%]. 
Also, it is worth pointing out that the share of respondents who never use ISO/IEC/IEEE 29148 [12;24.5\%] is larger than the share of respondents who never use its predecessor IEEE 830-1998 [8;16.0\%].
On the opposite side, two standards are intensively used by those respondents who know them, namely ISO/IEC/IEEE 12207 [24;80.0\%] and ISO/IEC/IEEE 15288 [21;80.8\%].

\subsubsection{RQ2.2: Purposes of standards used}

\begin{figure}
\centering
\includegraphics[width=\columnwidth]{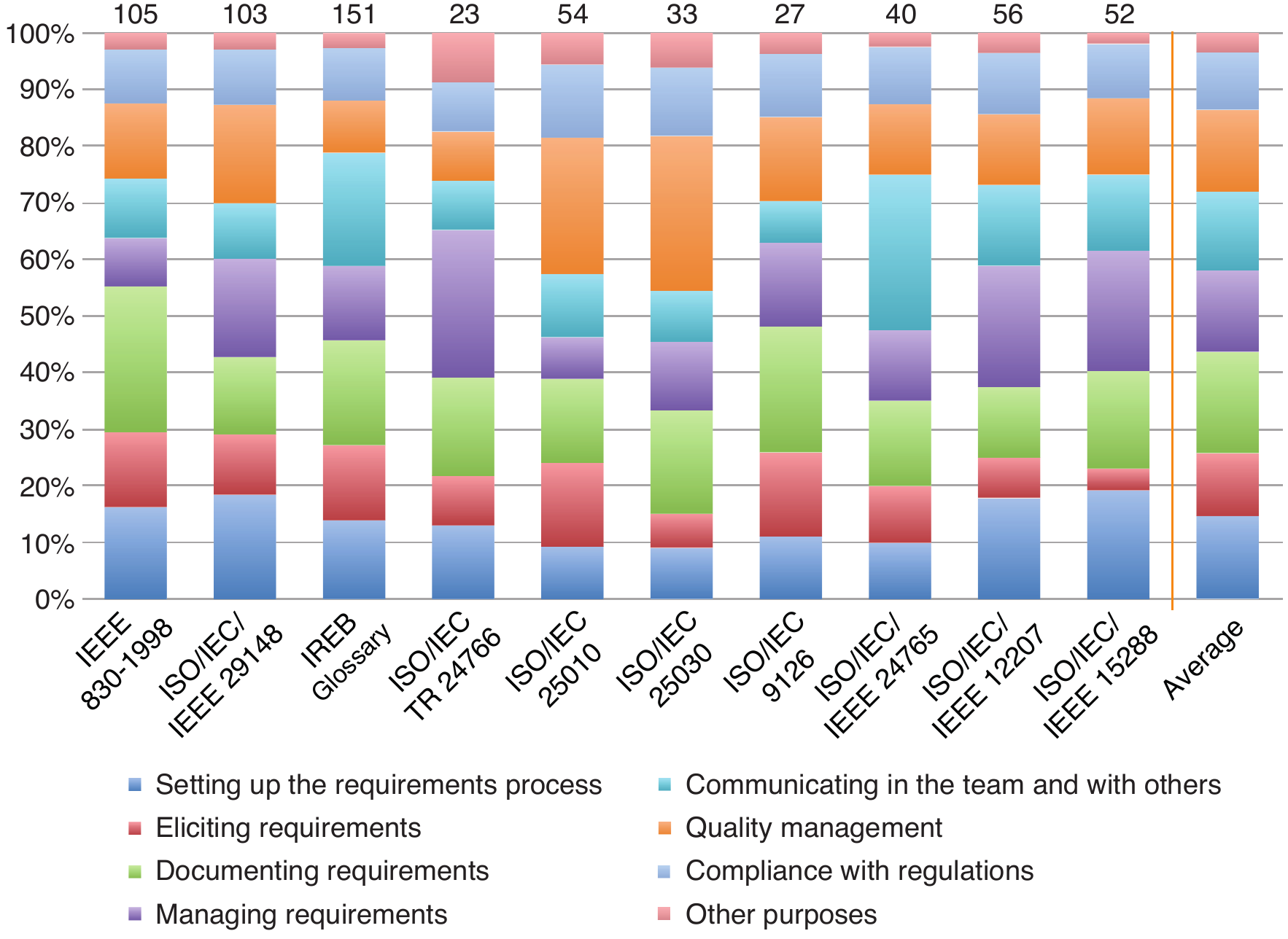}
\caption{RQ2.2: Purpose of using standards (number of responses given for each standard shown above the bars).}
\label{standards-purpose}
\end{figure}

When asking our respondents for the purposes they use standards for at least in some projects, 
we received a very mixed picture as shown in Fig.~\ref{standards-purpose}. A first observation is that in general, respondents use a single standard for several purposes that are reported in a relatively narrow band over all responses, from 17.9\% (Documenting requirements) down to 10.1\% (Compliance), with a 3.4\% of other purposes (e.g., ``Put in place governance around RE''). This narrow range shows that there is no single purpose clearly favored over the others when considering the ensemble of all RE-related standards and demands a closer look to individual standards.

Some standards' usage matches the original purpose of the standard. 
For IEEE 830-1998, for example, \emph{documentation} is the main purpose reported by the respondents [27;69.2\%], while \emph{management} is the least cited one [9;23.1\%]. This matches the standard's intention of providing a guideline for writing requirements specifications.
Also, the usage of the IREB Glossary primarily as a communication [30;65.2\%] and documentation [28;60.9\%] asset fits well with the intended purpose of a glossary. 

In contrast, there are standards whose usage does not match with the original purpose. 
For example, the quality standard ISO/IEC 9126 is not very often cited with the purpose of quality management [4;33.3\%]; instead, respondents report on documentation as its main purpose [6;50.0\%]. This may arguably be considered an unexpected usage. Conversely, its successor ISO/IEC 25010 is really considered for quality management [13; 59.1\%]. Other observations from our study are:
\begin{compactitem}
    \item All standards serve for all purposes to a certain extent. In fact, only two standards (ISO/IEC 25030:2007 and ISO/IEC/IEEE 15288:2015) rated a purpose below 15\%, and both of them referred to elicitation.
 \item The IREB Glossary and ISO/IEC/IEEE 29148 are the most versatile standards, with no purpose having a share of more than 20\% of responses.
 
    \item On the other hand, some standards are mentioned in relation to a main purpose. As mentioned earlier, IEEE 830-1998 is used primarily for documentation and ISO/IEC 25010 for quality management purposes. We also found that ISO/IEC TR24766 is mainly used for requirements management, ISO/IEC 25030 for quality management, and ISO/IEC/IEEE 24765 for communication. Nevertheless, the dominance is not big and there is no single purpose reaching the 30\% threshold over the number of respondents using the standard.
\end{compactitem}

\subsubsection{RQ2.3: Reasons for using standards}

\begin{figure}[t]
\centering
\includegraphics[width=\columnwidth]{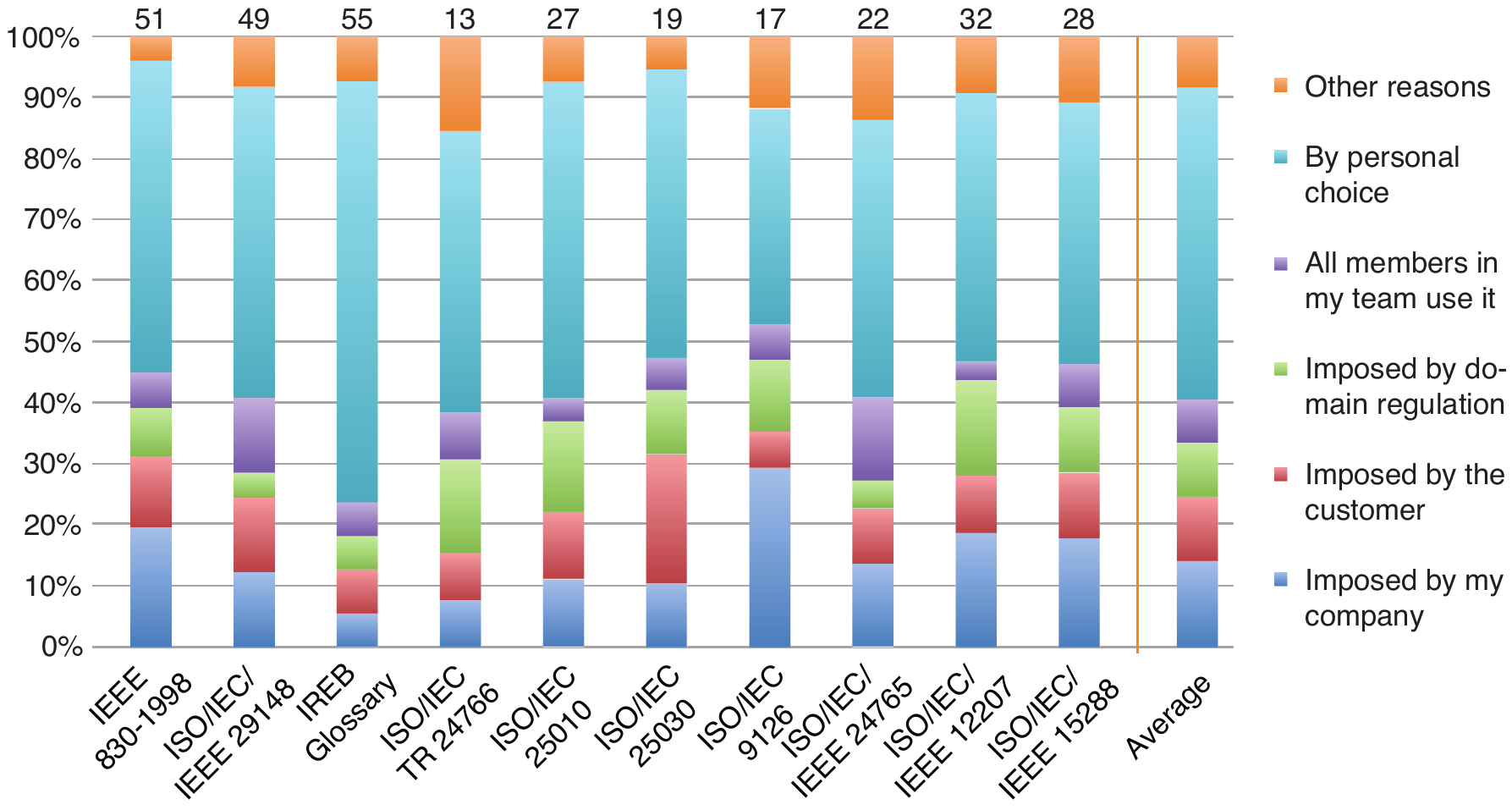}
\caption{RQ2.3: Reasons for using standards (number of responses given for each standard shown above the bars).}
\label{standards-reasons}
\end{figure}

We asked the respondents to state the reasons why they use the standards they know at least in some projects. Again, some respondents reported more than one reason per standard, but not so often as in the previous sub-question. 
In contrast to other questions in our survey, we received  a clear picture here (Fig.~\ref{standards-reasons}). The primary reason reported by respondents for using the standards is ``By personal choice''. On average, the respondents mentioned this reason more often than the other four reasons ``Imposed by the company", ``Imposed by the customer", ``Imposed by regulations" or ``Because all the team members use it'' combined.

This dominance is especially remarkable in the case of the IREB Glossary, where no other reason exceeds 10\%, compared to the [38;82.6\%] of personal choice. The only case in which this dominance is not so rampant is ISO/IEC 9126, where personal choice [6;50.0\%] is only slightly higher than company imposition [5;41.7\%]. 
Here it is also worth noting that an officially retired standard is imposed by so many companies.

\subsubsection{RQ2.4: Reasons for not using known standards}

\begin{figure}[!htb]
\centering
\includegraphics[width=\columnwidth]{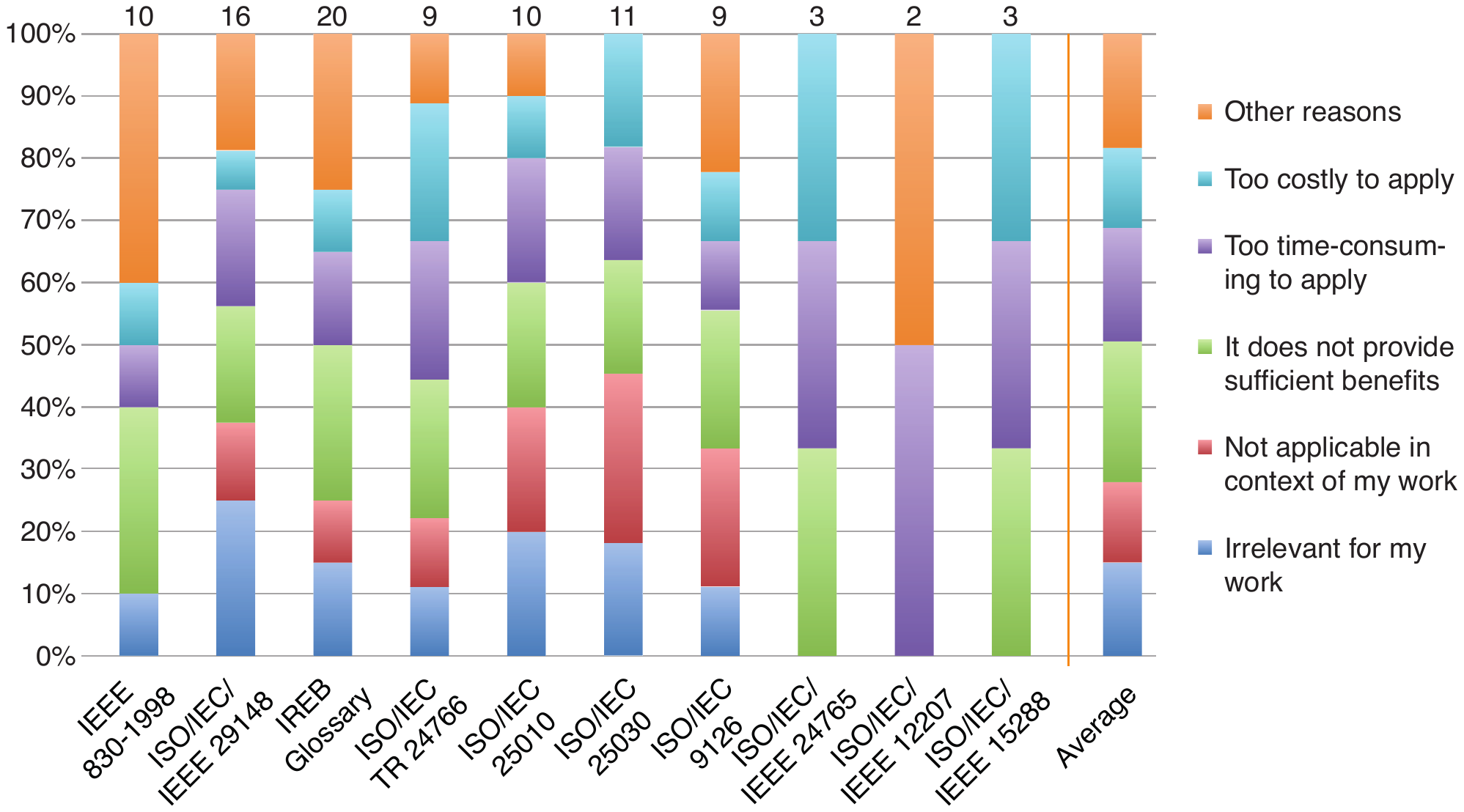}
\caption{RQ2.4: Reasons for not using standards known to our respondents (number of responses given for each standard shown above the bars).}
\label{reasons-not-using-standards}
\end{figure}

Fig.~\ref{reasons-not-using-standards} summarizes, for each of the known standards, reasons why our respondents don't use them at all. Overall, none of the reasons dominates, although not providing sufficient benefits is the most reported one (22.6\% over the total number of responses). It is interesting to note that cost and non-applicability are the least frequently mentioned reasons (12.9\% each). Among the other reasons provided as free text answers (18.3\%), we highlight ``Insufficient knowledge for a quick and efficient use", ``We haven't had the time to carefully revise the material", ``No need for now to apply them" and ``No specific guideline by my company". 

\subsection{RQ3. Under which conditions are RE-related standards used by practitioners?}\label{subsec:RQ3}
As mentioned in Sect.~\ref{sec:DataAn}, the answer to this RQ required the coding of three optional open-text fields, one per sub-question. Table \ref{tab:RQ3-codes} provides a summary. Responses are very diverse, ranging from one single word (``Regulation") to very elaborated sentences (one response up to 172 words mentioning five impediments in RQ3.2). Details on the answers to the RQs follow. 

\begin{table}[h]
\begin{threeparttable}
\caption{Coding overview for RQ3.}
    \label{tab:RQ3-codes}
    \centering
\renewcommand{\arraystretch}{1.3}
 \begin{tabular}{l | c c c c}
\emph{SubRQ} & \emph{Respondents\tnote{1}} & \emph{Terms\tnote{2}} & \emph{Topics} & \emph{Categories} \\
\hline
RQ3.1 & 38 & 64 (72) & 52 & 5 \\
RQ3.2 & 44 & 55 (78) & 28 & 4 \\
RQ3.3 & 45 & 57 (88) & 44 & 6 \\
\bottomrule
\end{tabular}
\begin{tablenotes}
\item[1] After filtering according to Sect.  \ref{sec:DataAn}
\item[2] Left, number of different terms. In parentheses, total number of references to these terms, considering repeated references in different responses.
\end{tablenotes}
\end{threeparttable}
\end{table}

\begin{table}[h]
\begin{threeparttable}
\caption{RQ3.1: Factors affecting standard adoption.}
    \label{tab:RQ3-1}
    \centering
\renewcommand{\arraystretch}{1.3}
\begin{tabular}{p{1.5cm} | p{6.2cm}}
\emph{Category\tnote{1}} & \emph{Factors\tnote{2}}  \\
\hline
Attribute & 6: Consistency, Quality \\
$\left[24;63.2\%\right]$ & 3: Adaptability, Completeness, Maturity \\
 & 2: Competence development, Efficiency, Satisfaction \\
  & 1: Abstraction, (Lack of) Ambiguity,  Clarity,  (Lack of) Complexity, Credibility,  Guidance,  Security, Testability \\
\hline
Activity  & 2: Audit, Communication, Compliance, \\
/Process & Understanding \\
$\left[11;28.9\%\right]$ & 1: (Required by) Bureaucracy, Certification, Governance, Quality assurance, Review, Training \\
\hline
Logical  & 4: (Best) Practices \\
Asset & 3: Body of Knowledge \\  $\left[11;28.9\%\right]$  & 2: Regulations, Vocabulary \\
& 1: Guidelines/Procedures, Reference framework  \\
\hline
RE Phases & 1: Elicitation, Management, NFR, Safety,\\
and Types $\left[5;13.2\%\right]$ & Regulatory \\
\hline
 Scope & 6: Requirement, External Stakeholder \\
$\left[27;71.1\%\right]$ & 5: Organization \\
 & 4: Product \\
 & 3: Market, Project, Team \\
 & 2: SRS \\
 & 1: Colleagues, Documentation, Industry (as a whole), IT infrastructure, Process, Self (the respondent), (Other) Standards \\
\bottomrule
\end{tabular}
\begin{tablenotes}
\item[1] [{\emph x};{\emph y}\%]: {\emph x}, number of respondents who mentioned factors in this category; {\emph y}\%, percentage regarding the 38 respondents.
\item[2] Digits show the number of responses citing every factor.
\end{tablenotes}
\end{threeparttable}
\end{table}
\subsubsection{RQ3.1. What factors affect the use of the standards?}
A slight majority of respondents [48;53.3\%] does not mention any factor driving the use of standards, while a few more [4;4.4\%] state that they do not know or give too vague or directly invalid statements.

We analyzed the responses given by the rest of the respondents ([38;42.3\%]) as explained in Sect.~\ref{sec:DataAn}. We identified five categories (see Table~\ref{tab:RQ3-1}):
\begin{compactitem}
\item \emph{Attribute.} Includes factors related to some quality aspect, e.g., consistency, quality or maturity. 
\item \emph{Activity/Process.} Identifies activities or processes that may benefit from the use of standards, as communication, audits or governance.
\item \emph{Logical Asset.} Enumerates concepts used in RE that may be improved through the use of standards, e.g., body of knowledge, vocabularies or regulations. 
\item \emph{RE Phases and Types.} Declares RE-specific phases and types in which some respondents recognize the need of applying standards, like requirements management and safety requirements.
\item \emph{Scope.} Binds the context of a given factor. Topics of this category act as modifiers of topics of the former.
\end{compactitem}

\begin{table}[ht]
\begin{threeparttable}
\caption{RQ3.2: Impediments on the effective use of standards.}
    \label{tab:RQ3-2}
    \centering
\renewcommand{\arraystretch}{1.3}
\begin{tabular}{p{1.5cm}  p{6.2cm}}
\emph{Category\tnote{1}} & \emph{Factors\tnote{2}}  \\
\hline
Knowledge & 9: Insufficient knowledge/skills \\
-related & 6: Company capacity \\
$\left[19;43.2\%\right]$ & 3: Lack of awareness  \\
 & 2: Application/Project domain, Lack of experts \\
  & 1: Company context, Limited industrial adoption   \\
\hline
Intrinsic & 3:  Difficult to read, Not useful, Rigid,  \\
 $\left[16;36.4\%\right]$ & Too theoretical \\
 &  1: Copyright limitations, Resources to invest, Too abstract,   Too complicated, Too many standards, Too simple \\
\hline
Cultural & 8: Poor consideration of RE in company \\
 $\left[14;31.8\%\right]$ & 6: Employees' attitude \\
 & 4: Company culture \\
 & 1: Processes considered unnecessary, Quality not sought, Resistance to standards  \\
 \hline
Operational & 4: Poor effectiveness \\
$\left[10;22.7\%\right]$ & 3: Need of tailoring \\
 & 2: Cost pressure \\
  & 1: Lack of tool support, Time pressure  \\
\bottomrule
\end{tabular}
\begin{tablenotes}
\item[1]  [{\emph x};{\emph y}\%]: {\emph x}, number of respondents who mentioned factors in this category; {\emph y}\%, percentage given over the 44 respondents who provided valid answers.
\item[2] Digits show the number of responses citing every factor.
\end{tablenotes}
\end{threeparttable}
\end{table}

\subsubsection{RQ3.2. What are the major impediments in effectively using RE-related standards?}
Almost half of the respondents [43;47.8\%] do not mention any impediment regarding the use of standards, while a few more [3;3.3\%] state that they are not aware of any or provide an invalid statement.

We grouped the impediments emerging from the responses given by the [44;48.9\%] remaining respondents into four categories (see Table \ref{tab:RQ3-2}):
\begin{compactitem}
\item \emph{Knowledge-related.} Includes factors related to the knowledge or mastering of standards, e.g., lack of awareness of their existence or shortage of resources to invest in their acquisition or application. 
\item \emph{Intrinsic.} Enumerates characteristics that the respondents assign to standards, e.g., they are too theoretical, too complicated or simply not useful.
\item \emph{Cultural.} Collects some factors that act as impediments due to the culture of the organization, for example employees' lack of motivation or the under-estimation of RE by managers. 
\item \emph{Operational.} Declares barriers that emerge once the standard has started to be applied, like time pressures or the need of tailoring to the specific needs of the company or project.
\end{compactitem}

\begin{table}[hbt!]
\begin{threeparttable}
\caption{RQ3.3: Perceived benefits from the use of standards.}
    \label{tab:RQ3-3}
    \centering
\renewcommand{\arraystretch}{1.3}
\begin{tabular}{p{1.5cm}  p{6.2cm}}
\emph{Category\tnote{1}} & \emph{Factors\tnote{2}}  \\
\hline
Attribute & 11: Quality \\
$\left[20;44.4\%\right]$ & 4: Consistency \\
 & 3: Clarity, Completeness  \\
 & 2: Compliance, Effectiveness \\ 
 & 1: Credibility, Timeliness (state-of-the-art) \\
\hline
Methodology & 4: Evolution \\
and IT & 3: Approach to development \\
 $\left[15;33.3\%\right]$ & 2: Acceptance of results, Reuse, Structure (of process and documentation) \\
 & 1: Design \& Implementation, Good practices, Prevision, Project handling, Technology leverage,  \\
\hline
Human & 4: Common understanding, Communication \\
Factors & 3: Agreed terminology, Shared language,  \\
$\left[18;40.0\%\right]$ & (Support to) Argumentation \\
 & 1: Established knowledge, Giving advice,  Professional approach, Rationale (for decisions), (Personal) Satisfaction \\
\hline
Project & 3: Cost/Budget, Risk  \\
Factors & 2: Time \\
$\left[6;13.3\%\right]$ & 1: Effort
 \\
\hline
Organiza- & 2: Customer satisfaction \\
tional & 1: Finances, Team management, \\
$\left[5;11.1\%\right]$ & Awareness on RE \\
\hline
RE- & 2: Specification \\
related & 1: Body of knowledge, Business analysis, \\
$\left[7;15.6\%\right]$ & Business alignment, Elicitation, Glossary, Management  \\
\bottomrule
\end{tabular}
\begin{tablenotes}
\item[1]  [{\emph x};{\emph y}\%]: {\emph x}, number of respondents who mentioned factors in this category; {\emph y}\%, percentage given over the 45 respondents who provided valid answers.
\item[2] Digits show the number of responses citing every factor.
\end{tablenotes}
\end{threeparttable}
\end{table}

\subsubsection{RQ3.3. What are the perceived advantages coming from the use of RE-related standards?}
Nearly half of the respondents [42;46.7\%] do not mention any perceived advantage from using standards, and a few [3;3.3\%] state that they are not aware of any or provide confounding information.

We proceeded similarly as for the previous sub-questions and grouped the benefits emerging from the rest of responses [45;50.0\%] into six categories (see Table \ref{tab:RQ3-3}):
\begin{compactitem}
\item \emph{Attribute.} Includes factors directly related to improved properties of the requirements, e.g., their quality, consistency, clarity or completeness. 
\item \emph{Methodology and IT.} Enumerates aspects of the method or the technological solution which may benefit from standards, for instance, easier product evolution or improved project management.
\item \emph{Human factors.} Enumerates improved human abilities or skills, e.g., providing better arguments in decision-making or better communication abilities.
\item \emph{Projects factors.} Collects some factors that are related to the project measurement or success, like risk, time or effort. 
\item \emph{Organizational.} Declares benefits directly perceived by the organization, as customer satisfaction or team management.
\item \emph{RE-related.} Identifies benefits related with the RE process, like improved requirements elicitation or management, or a more comprehensive body of knowledge (``structure to the [RE] field").
\end{compactitem}

 \subsection{RQ4. Do practitioners use further guidelines or documentation templates in their RE process?}\label{subsec:RQ4}

 RQ4 includes four research sub-questions. Answers to RQ4.2-RQ4.4 rely on the topics emerging from RQ4.1.
 
 \subsubsection{RQ4.1. What further guidelines or templates do practitioners use?}
\begin{figure}[htb]
\centering
\includegraphics[width=0.5\textwidth]{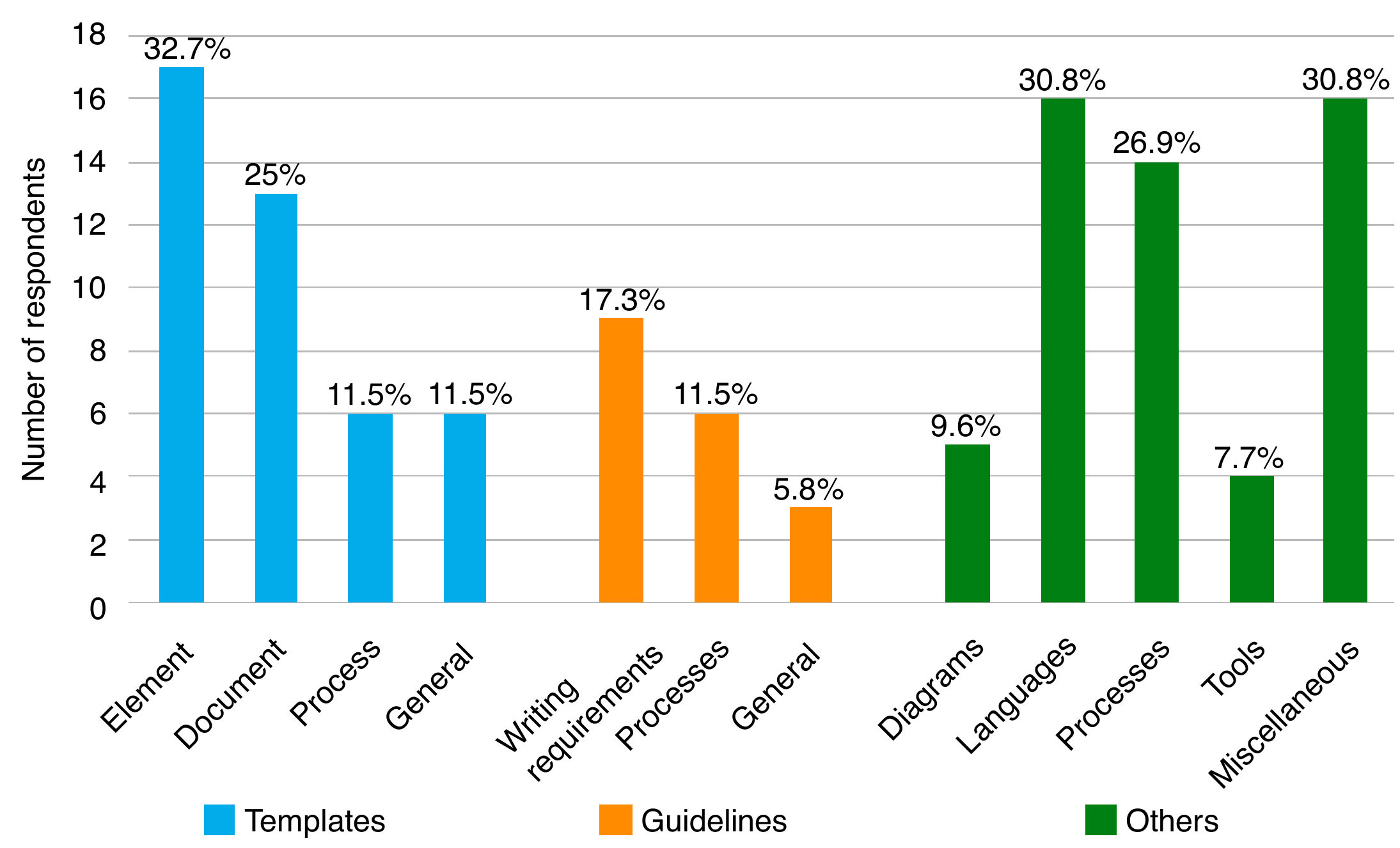}
\caption{RQ4.1: Overview of respondents' use of templates and guidelines (percentages given over the set of 52 respondents who provided valid responses).}
\label{RQ4_1_overview}
\end{figure}

 We applied the analysis method described in Section \ref{sec:DataAn} over a total of 136 responses from 54 respondents. Ten responses had to be discarded because they referred to standards instead of guidelines and templates (and should have been provided as responses to RQ1.1 accordingly). One respondent explicitly stated not to use any guidelines or templates. As a result, we were left with 125 responses from 52 respondents (i.e., 57,8\% over the total number of respondents). Some responses referred to more than one concept and after decomposing them, we finally came up with 149 individual references to guidelines, templates and other standardized artifacts. 
We identified 65 primary codes to classify these responses. For 28 of them, we identified a secondary code providing further information (e.g., a particular tool or a particular technique). In total, we identified 34  secondary codes.
 However, the analysis of the secondary codes did not yield any notable insights with respect to RQ4.1. The primary codes were grouped into three categories following the question made, further divided into subcategories (see Table~\ref{tab:RQ4-1} for the coding and Fig.~\ref{RQ4_1_overview} for the number and percentage of respondents using every type of artifact):
 \begin{itemize}
     \item \emph{Templates}. Writing templates are the dominant type, both for individual elements (e.g., use cases, user stories) and entire documents (typically, the requirements specification). Some respondents reported templates for RE-related processes (e.g., product gap analysis), while a few more just reported the use of templates in general or did not provide details.
     \item \emph{Guidelines}. We followed a similar sub-classification as above, although in this case we didn't distinguish among guidelines for writing individual elements or documents because the responses were in general not very specific regarding this aspect.
     \item \emph{Others}. Here we grouped all the codes corresponding to responses that were neither templates nor guidelines. Mainly, modeling languages, processes other than RE-related ones, tools, diagrams, and other miscellaneous (e.g., traceability matrices, ontologies, architecture development methods).
 \end{itemize}


\begin{table*}[ht]
\begin{threeparttable}
\caption{RQ4.1: Overview of responses and respondents concerning the use of templates and guidelines.}
    \label{tab:RQ4-1}
    \centering \footnotesize
\renewcommand{\arraystretch}{1.3}
\begin{tabular}{p{1.4cm}| p{4.8
cm} | p{10cm}}
\emph{Category\tnote{1}} & \emph{Sub-category\tnote{1}} & Primary Code (examples)\tnote{2} \\
\toprule
\multirow{4}{=}{Templates [34;65.4\%]} &  Element level template [17;32.7\%] & Functional requirements template, User story \\ \cline{2-3}
 & Document level template [13;25.0\%] & Business requirements documentation template, System requirements specification template \\ \cline{2-3}
 & Process support template [6;11.5\%] & Change management template, Scoping template \\ \cline{2-3}
 & Templates in general [6;11.5\%] & General templates, Requirements documentation templates \\ \hline
  \multirow{3}{=}{Guidelines [14;26,9\%]} &  Guidelines for writing requirements [9;17.3\%] & Requirements writing guideline, Specification writing guideline \\ \cline{2-3}
 & Guidelines for requirements-related processes [6;11.5\%] & Configuration management guideline, RE process guideline \\ \cline{2-3}
 & Guidelines in general [3;5.8\%] & General guidelines \\ \hline
  \multirow{5}{=}{Others [34;65.4\%]} &  Standardized diagrams [5;9.6\%] & Activity diagram, Class diagram \\ \cline{2-3}
 & Standardized languages [16;30.8\%] & Modelling language \\ \cline{2-3}
 & Standardized processes [14;26.9\%] & Change request process, Software development process \\ \cline{2-3}
 & Tools [4;7.7\%] & Tool \\ \cline{2-3}
 & Miscellaneous other [16;30.8\%] & Architecture development method, Body of knowledge, Data dictionary,  Product vision board, Test case \\ \bottomrule
\end{tabular}
\begin{tablenotes}
\item[1] In [{\emph x};{\emph y}\%], {\emph x} is the number of respondents who mentioned primary codes in this category, and {\emph y}\% is the percentage of {\emph x} over the number of respondents who use standardized artifacts (52).
\item[2] For the full list of codes, we refer to the study data package available at \url{http://doi.org/10.5281/zenodo.4755756}.

\end{tablenotes}
\end{threeparttable}
\end{table*}

The number of respondents reporting use of templates more than doubles the number related to the use of guidelines. Furthermore, the number of respondents who reported the use of both templates and guidelines is as low as [10;19,2\%]. 
 The number of respondents referring to other artifacts is the same as the number of respondents using templates. Most usual references were to modeling languages (especially UML) and processes (remarkably Scrum). We also observed that respondents who use other artifacts tend to use several of them together, e.g., various types of UML diagrams.

 \subsubsection{RQ4.2. How often are these guidelines or templates used?}

Table \ref{tab:RQ4.2-frequency} shows the frequency of use of templates, guidelines and other standardized artifacts. All categories have a similar distribution, with close to half of the responses reporting use in every project and with all categories around 20\% of scarce use (``in some project"). 

We have also analyzed usage at the level of the primary codes and in general differences are not remarkable. We only remark the following cases:
\begin{itemize}
    \item The usage of guidelines for writing requirements is highly polarized, with 4 usages out of 10 in every project, and 5 other usages only in some project.
    \item On the contrary, the usage of diagrams concentrates in the mid-range of usage (``in most projects"), as stated in 12 out of the 13 reported usages.
    \item All 5 usages of tools apply to every project.
\end{itemize}
 
 \begin{table}[h]
\begin{threeparttable}
\caption{Summary of use of templates, guidelines and other standardized artifacts ({\it Usage} refers to projects).}
    \label{tab:RQ4.2-frequency}
    \centering \footnotesize
\renewcommand{\arraystretch}{1.3}
\begin{tabular}{p{1.3cm} | p{1.1cm} p{1.1cm} p{1.1cm} p{1.1cm}}

\emph{Usage}\tnote{1} & \emph{Templates} & \emph{Guidelines} & \emph{Others} & \emph{Overall} \\
\hline
Every  & [25;50.0\%] & [10;45.5\%]  & [33;42.9\%] & [68;45.6\%] \\
Most  & [15;30.0\%] & [8;36.4\%]  & [28;36.4\%] & [51;34.2\%] \\
Some  & [10;20.0\%] & [4;18.2\%]  & [16;20.85\%] & [30;20.1\%] \\
\hline
Total & 50 & 22  & 77 & 149 \\\bottomrule
\end{tabular}
\begin{tablenotes}
\item[1] [{\emph x};{\emph y}\%]: {\emph x}, number of responses that report the given level of use for the given category; {\emph y}\%, percentage of {\emph x} over the number of responses for the given category (which appears in the last row).
\end{tablenotes}
\end{threeparttable}
\end{table}
 
\subsubsection{RQ4.3. For which purpose do practitioners use each of these guidelines or templates?}
Table \ref{tab:RQ4.3-purposes} shows the frequency of the purposes of use of templates, guidelines and other standardized artifacts. Documentation of requirements is by far the main purpose of use, especially in the case of templates. The rest of purposes do not reach 17\% of the responses given. Interesting results are the high percentage of responses reporting communication purposes for artifacts other than templates and guidelines, as well as the relative importance of guidelines for requirements management and setting up the RE process. Most of the respondents who mentioned ``Other" simply wanted to express multiple purposes for an artifact.

As in the previous sub-question, we analyzed the individual artifacts in detail. We found the following:
\begin{itemize}
    \item Although documentation is the main purpose of both document level templates and element level templates, the dominance is different. While at the document level it is the dominant purpose (12 out of 14), only 9 out of 21 mention this purpose on the element level. 
    \item The main purpose of the different types of other artifacts varies: diagrams for eliciting requirements [7;53.8\%], languages for documenting them [10;52.6\%], processes for communication purposes [7;43.8\%] and tools for managing requirements [2;40.0\%]. 
\end{itemize}

\begin{table}[htb]
\begin{threeparttable}
\caption{Summary of purposes of using templates, guidelines and other standardized artifacts.}
    \label{tab:RQ4.3-purposes}
    \centering \footnotesize
\renewcommand{\arraystretch}{1.3}

\begin{tabular}{p{1.3cm} | p{1.1cm} p{1.1cm} p{1.1cm} p{1.1cm}}


\emph{Usage}\tnote{1} & \emph{Templates} & \emph{Guidelines} & \emph{Others} & \emph{Overall} \\
\hline
Docu-mentation & [23;46.0\%] & [9;40.9\%]  & [22;28.6\%] & [54;36.2\%] \\
Commu-nication & [5;10.0\%] & [0;0.0\%]  & [20;26.0\%] & [25;16.8\%] \\
Elicita-tion & [5;10.0\%] & [1;4.5\%]  & [11;14.3\%] & [17;11.4\%] \\
Mana-gement & [5;10.0\%] & [4;18.2\%]  & [7;9.1\%] & [16;10.7\%] \\
Set up process & [3;6.0\%] & [4;18.2\%]  & [6;7,8\%] & [13;8.7\%] \\
Quality mgmt. & [3;6.0\%] & [3;13.6\%]  & [3;3.9\%] & [9;6.0\%] \\
Com-pliance & [3;6.0\%] & [1;4.5\%]  & [1;1.3\%] & [5;3.4\%] \\
Others & [3;6.0\%] & [0;0.0\%] & [7;9.1\%] & [10;6.7\%] \\
\hline
Total & 50 & 22  & 77 & 149 \\\bottomrule
\end{tabular}
\begin{tablenotes}
\item[1] [{\emph x};{\emph y}\%]: {\emph x}, number of responses that report the given level of use for the given category; {\emph y}\%, percentage of {\emph x} over the number of responses for the given category (which appears in the last row).
\end{tablenotes}
\end{threeparttable}
\end{table}

 \subsubsection{RQ4.4. How are these guidelines or templates used in combination with RE-related standards?}
Up to 56 respondents provided 103 responses to this last sub-question (which includes two respondents who did not provide answers to the previous sub-questions). This means that some respondents (concretely, 33) reported more than one type of combination; unfortunately, the questionnaire did not gather explicit information about the context of each type, and the respondents did not use the free text to inform either. 

\begin{figure}[!htb]
\centering
\includegraphics[width=\columnwidth]{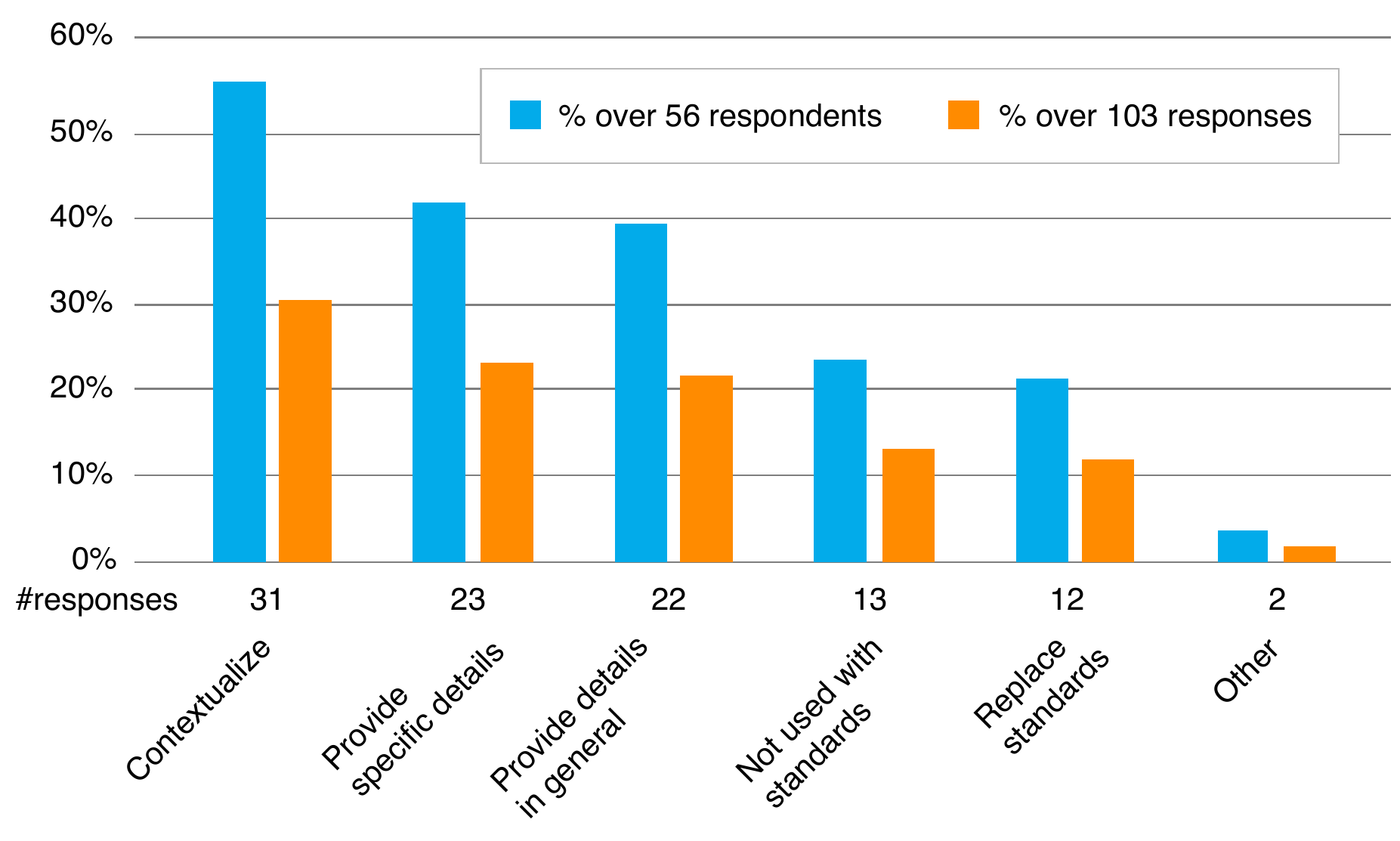}
\caption{RQ4.4: Uses of standardized artifacts with standards.}
\label{art-with-st}
\end{figure}

Fig. \ref{art-with-st} shows that the greatest percentage of respondents reported use of standardized artifacts as a way to add detail to standards in three possible different ways (provide context, add specific details, or add details in general), while the number of respondents using the artifacts as replacement of, or independently of, standards did not reach 25\% in any case.

~

~

\section{Discussion}\label{sec:analysis}
In this section, we discuss our major findings in detail. 

\emph{Finding 1: The general knowledge about RE standards is arguably rather low.} As shown in Sect.~\ref{RQ1_1}, 42.2\% of the respondents do not know more than two RE-related standards. We found similar percentages for the respondents who do not know the core RE standard, ISO/IEC/IEEE 29148, and its predecessor IEEE 830-1998: 45.6 percent and 44.4 percent, respectively. 
If we analyze individual responses in detail, 27.8\% of all respondents   
do not know any of these two core standards
. Even the most known standard in our study, the IREB Glossary of RE Terminology, is not known by 30\% of the respondents, and this number grows to 51,1\% if we consider those respondents without any IREB certification.


In summary, and this is a main result of our study, we found that \emph{the level of knowledge of RE-related standards is not as high as one could expect}. This lack of knowledge may prevent practitioners from making informed decisions in their projects, for example, whether to apply a certain method, follow a certain template or use the right terminology. This does not mean that knowing about standards will always lead to their application, but knowing a standard allows to decide against its application based on well-justified reasons.

\begin{figure}[!t]
\centering
\includegraphics[width=\columnwidth]{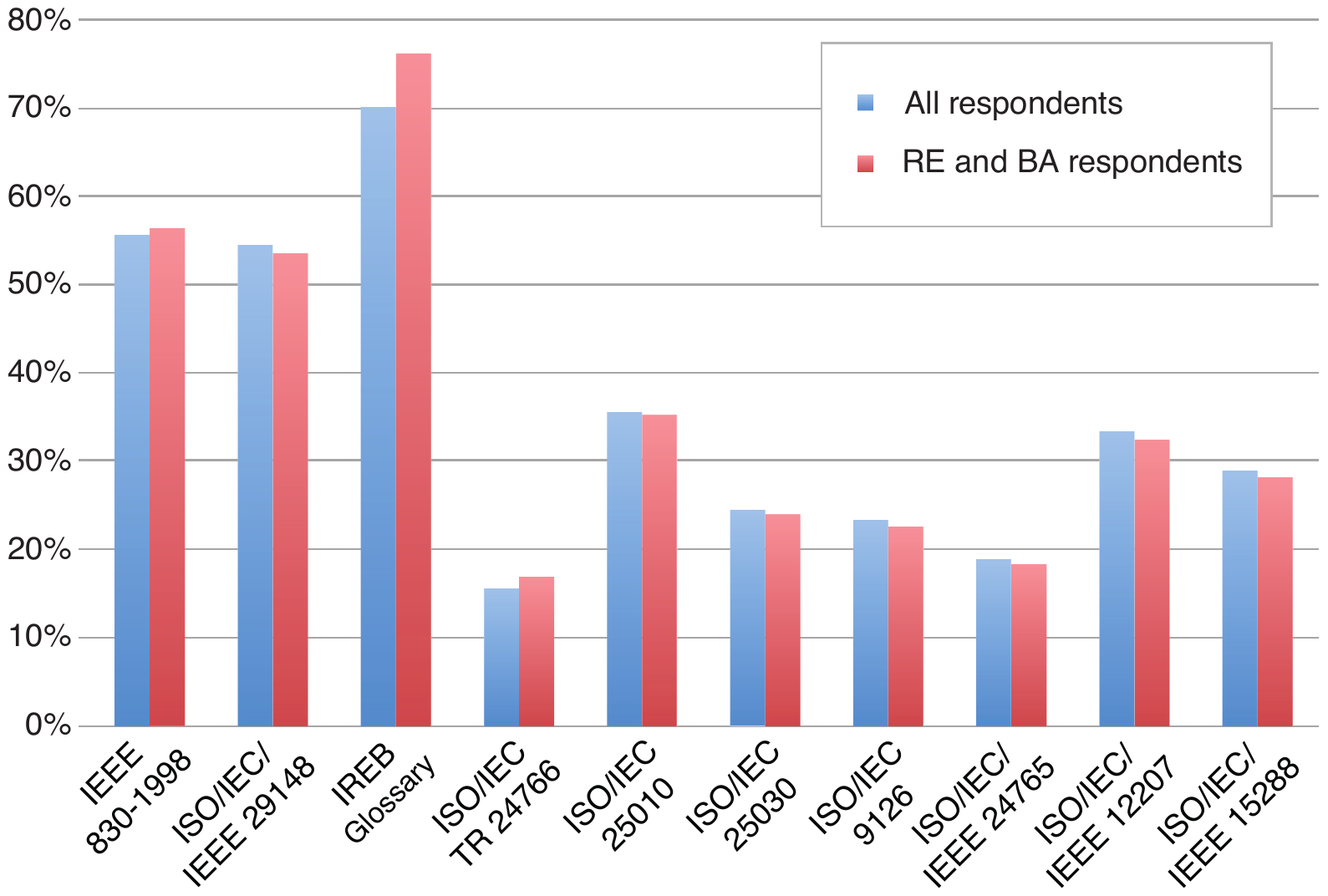}
\caption{Knowledge of RE-related standards by requirements engineers and business analysts.}
\label{fig:standards-known-BAs}
\end{figure}

We discuss next if there is any demographic factor with an influence on the knowledge of standards. Remarkably, one could expect that the 71 respondents who hold or held the job title {\it requirements engineer} or {\it business analyst} in the past six years would be more knowledgeable. However, as Fig.~\ref{fig:standards-known-BAs} shows, differences are not important. In the same vein, the knowledge of quality standards by the 18 respondents occupying positions of quality engineers/quality managers is not massive, with 61.1\% of them knowing ISO/IEC 25010 and only 38.9\% and 33.3\% knowing ISO/IEC 25030 and ISO/IEC 9126, respectively. Therefore, we conclude that according to our study, \emph{the job title is not determinant for the knowledge of standards}.

We also analyzed the possible relation with the experience level. In general, results suggest that more experienced professionals have a slightly wider knowledge of standards. For instance, the percentage of respondents with more than ten years of experience who know IEEE 830-1998 or ISO/IEC/IEEE 29148 is 82.1\%, compared to the 64.7\% of the rest of the population. We found similar results for other standards that suggest that \emph{experience has an influence on the knowledge of standards}. 

We also compared knowledge of standards with the class of systems developed by our respondents. Although most of the knowledge was evenly distributed, we found that the 20 respondents working on embedded systems have scarce knowledge of quality-related standards (a total of [16;80\%] do not know any) but instead are more aware of other engineering standards ([12;60\%] know at least one), which is opposite to what we found for the rest of system classes.  
Therefore, \emph{system class may be determinant in the type of standards that practitioners need to master for their work}.

We analyzed other possible influencing factors such as size or role of the team, but couldn't find any further relationships, except certification. Our general observation is that only 3.8\% of certified respondents do not know any standards, in contrast to the 13.2\% of non-certified respondents. This indicates a \emph{positive influence of certification on standard knowledge}. However, some particular standards do not follow this general trend, and, in particular, the knowledge of IEEE 830-1998 or ISO/IEC/IEEE 29148 is slightly lower for certified respondents than for non-certified respondents (71.2\% vs. 73.7\%), with the only exception of the advanced level certification population segment, for which the percentage increases to 84.6\%. 


\emph{Finding 2: Even the known standards are barely used.}
Even if respondents know the standards, they still barely use them. For example, only 14.3\% of the respondents who know ISO/IEC/IEEE 29148 actually use it in all their projects, while almost two thirds of them (65.3\%) use it only in some projects or even never.
Again, the situation does not improve much when considering the job title. For instance, when combining the numbers for not knowing or never using standards, [26;36.6\%] of the RE/BA respondents do not know or never use the two core RE standards, ISO/IEC/IEEE 29148 and its predecessor IEEE 830-1998.

The numbers are similarly striking if we focus on respondents instead of standards: 37 out of the 83 respondents who declared to know some standard (i.e., 44.6\%) use \emph{all of them} scarcely (i.e., they use all standards they know 
only in some projects or never, or they prefer not to respond\footnote{Arguably, we consider declining of a response as an indicator of scarce use.}).

Similarly to above, we investigated if some demographic factors could have an impact on the use of standards. We focused again on experience level and certification. While we did not find any  relationship among level of experience and use of standards, we found that respondents without any certification are more prone to use standards scarcely than the rest (54.5\% compared to 38.0\% for certified respondents). Considering this observation and also the increased level of general knowledge, we may conclude that \emph{certification has a positive influence on RE-related standards knowledge and use}.

There may be further reasons behind the limited use of standards. For instance, as early as in 1994, Pfleeger et al. remarked as criticism at that time that ``standards have codified approaches whose effectiveness has not been rigorously and scientifically demonstrated” \cite{Pfleeger94}. Another reason can be the time needed to develop and deliver a standard. Phillips~\cite{Phillips2019} found that it typically takes three years to create an ISO standard. Therefore, the contents of a standard may already look a bit outdated when the standard is finally published. Last, in software engineering in general, Laporte et al.~\cite{Laporte18} have reported as reasons for not applying standards the perception that they are meant for larger organizations and require much documentation and bureaucracy.

\emph{Finding 3: Superseded old standards are still in use.}
Older standards, such as the IEEE 830-1998, are still being used as a documentation guideline despite being superseded by follow-up standards. One reason could be that IEEE 830-1998 is much shorter and conciser than ISO/IEC/IEEE 29148. Since our results also show that documentation is the most cited purpose for standards in general, it might be even the case that IEEE 830-1998, which is especially appreciated by respondents in this regard, could be hard to replace completely by ISO/IEC/IEEE 29148.

Other reasons could also apply. For instance, due to the lack of continuous training and learning, practitioners might not gain awareness about new standards and superseded old ones. In fact, only two respondents mentioned at some point that the former standards were superseded by the new ones, which supports this argument. Also, in RQ2.3, we could observe that the two officially retired standards IEEE 830-1998 and ISO/IEC 9126 are those with the highest percentages of being imposed by organizations. This could be linked with some of the impediments reported in RQ3.2, as the lack of knowledge in the organization. Other possible reasons could be reluctance to change, especially if practitioners cannot perceive a real benefit in ISO/IEC/IEEE 29148 compared to IEEE 830-1998, or that the communication of a new standard when it is released is not effective. We couldn't find any further evidence from the data. For instance, we explored if respondents with more experience were not aware of the new standards and stuck to the old ones, but this was not the case.

\emph{Finding 4: University curricula do not communicate the existence of standards effectively}. Our results in Fig.~\ref{fig:standards-sources} show that even if the great majority of our respondents have university degrees, university studies take only the 5th place when it comes to where the respondents learned about the standards, with only [18;22.1\%] of respondents citing university studies as a source. 
Only respondents with a PhD yield a greater share of up to 33.3\%, but this segment of the population is small (only 9 respondents in our study). For the rest of respondents, and especially those holding an engineering degree, training courses and self-learning (reading or web searches) were much more influential.

This observation collides with some criteria formulated by the Accreditation Board for Engineering and Technology (ABET), e.g., ``students must be prepared for engineering practice through a curriculum culminating in a major design experience based on the knowledge and skills acquired in earlier course work and incorporating appropriate engineering standards and multiple realistic constraints” \cite{ABET2017} and suggests the need for RE educators in university settings to give more attention to this area. This action has to overcome reported impediments to teaching standards as lack of text books, lack of practical knowledge by educators, and availability and cost of access to technical standard documents \cite{KKC13}, \cite{Phillips2019}. Innovative approaches such as gamification \cite{Garcia20} may help educators to improve the teaching of standards.

\emph{Finding 5: Standards do not provide adequate support neither for requirements elicitation nor for compliance with regulations.}
Our results show that these two aspects are the worst rated purposes for which standards are used with averages of 11.0\% and 10.1\%, respectively, of all responses (see Fig.~\ref{standards-purpose}). When looking at respondents instead of responses, we get a similar picture: elicitation and compliance are the worst rated purposes, with less than 30 percent of respondents using standards for these purposes. 
Especially noticeable is the case of requirements elicitation, given that this activity is recognized as crucial in the RE discipline. In our opinion, this finding calls for an effort by standard agencies and organizations to improve this aspect in later versions of current standards, or even new standards.


\emph{Finding 6: There is a great variety of factors driving the use of standards.}
The dominant reason for using standards is \emph{Personal Choice} (see Fig.~\ref{standards-reasons}), with an average of 51.1\% and a range from 35.3\% for ISO/IEC 9126 to 69.1\% for the IREB Glossary. When taking together the cases where standards are imposed by regulators, customers or the respondents' organization, \emph{being imposed} is the follow-up reason for using standards, ranging from 18.2\% for the IREB Glossary to 47.1\% for ISO/IEC 9126 and an average of 33.5\%. The former is no surprise, as the IREB Glossary is no official standard, while the latter aligns with {\it Finding 3}, as ISO/IEC 9126 was officially withdrawn in 2011, when its successor, ISO/IEC 25010 was released.

As the results for RQ3 (see Sect.~\ref{subsec:RQ3}) show, respondents provide many reasons for justifying the use of standards. A high share (63.2\%) mention specific factors as drivers on the use of standards, and among them, consistency and quality are the most frequently cited ones. As mentioned in one response, ``when dealing in large teams, diverse processes and workshops running in parallel, the standards help in ensuring consistency on the level of details documented to help in smoother implementations". Communication at all levels is also a frequently cited factor, as reflected, e.g., by ``Common understanding in the team", ``Ensure a common understanding of the requirements with internal and external stakeholders", and ``To introduce a common RE language in my department (RE glossary)".

\emph{Finding 7: Lack of knowledge and skills, and personal and organizational culture, are the main impediments to the adoption of standards.}
The greatest share of respondents reporting impediments ([33;75.0\%]) identify different knowledge-related or cultural issues (see Table~\ref{tab:RQ3-2}\footnote{Please take into account that a respondent may identify several factors as impediments.}). Resolving them could  pave the way to improved standards adoption.

Cultural issues are the most striking ones since they seem difficult to be solved in an organization in the short term. Among them, the poor consideration of RE (8 responses) goes beyond the standards adoption problem. A respondent summarized the problem as ``Too [few] people know about the challenge of the work with requirements. Project managers think you just ask the business and that's it". Moreover, from a business point of view, ``most companies lack [an] RE process because they don't see the value". Another respondent stated  ``[A] realistic budget is almost never allocated to RE"
In this context, it is not just the adoption of RE-related standards that is jeopardized, but the whole RE process. 

Last, the percentually high number of respondents who complain about  standards {\it per se} (16 respondents, i.e. 36.4\% of those who mentioned impediments) cannot be neglected. Some criticisms look very well justified and should be considered as useful input for standard organizations. For example, standards are said to be too rigid, too theoretical, or to put it simpler: ``they do not answer the real needs". 
However, other responses express beliefs that show a misconception about how current standards may be used such as ``Our environment is changing too fast to use an old fashioned standard" or ``our standard way of specifying requirements is using natural language, so we cannot benefit from modelling standards". This points out the \emph{need of a better communication of the utility of standards}.

\emph{Finding 8: Human factors do equally benefit from the use of RE-related standards as the quality of requirements.}
More than a third of our respondents who reported benefits ([18;40.0\%]) mentioned that human-related skills or activities have improved after the use of RE-related standards.
This is equal to the share of respondents mentioning benefits related to an improvement of the quality\footnote{Respondents mentioning the attributes \emph{consistency}, \emph{quality}, \emph{compliance}, \emph{completeness}, \emph{credibility}, or \emph{clarity}; see Table~\ref{tab:RQ3-3}. We omitted the attributes \emph{effectiveness} and \emph{timeliness} as they refer to process quality.} of the requirements ([18;40.0\%]) and underlines that the importance of standards transcends a purely technical perspective on RE.

From a communication perspective, all involved stakeholders benefit from standards: ``[there is a] common understanding among the people working on requirements'', ``easier communications with customers and other counter-parties''. ``Glossary standardization'' is one of the referenced assets supporting communication.


Another emerging human-related aspect is the confidence that standards give to requirement engineers to justify their decisions, e.g., ``the BA always has the necessary tool set for arguments'' or ``I have been challenged in the past for why I have chosen to adopt a certain methodology or approach and having a standard to reference has allowed me to promote my opinion more effectively''. Being aware of this may help requirements engineers realizing about personal outcomes they may take from standards' adoption.

Needless to say, the impact on human factors does not mean that technical benefits should be dismissed. It is not just ``higher quality of requirements'', but as a consequence ``higher project success and final product quality; improved customer satisfaction''. One of the respondents even provided a rough estimate of this gain: ``Up to 30\% less time and budget consuming for entire project effort (not at the beginning, but over time)''. 

\emph{Finding 9: The use of standards and standardized artifacts appears to be independent.}
When analyzing the usage of standardized artifacts, it could be expected that there is some kind of relationship with the usage of standards. We investigated two particular hypotheses:
\begin{itemize}
    \item(1)~Respondents knowing less standards, or using less often the standards that they know, use more standardized artifacts, or use them more often. This would be an indicator that standardized artifacts may be used as a kind of substitute for standards.
    \item(2)~Respondents knowing more standards, or using more often the standards that they know, use more standardized artifacts, or use them more often. This would be an indicator that the profile, attitude or education of the respondent is more determinant than the type of resource used.
\end{itemize}

However, we couldn't find any relationship among these concepts, which lead us to conclude that according to this study, \emph{the use of standards and other artifacts as templates, guidelines, notations and languages, are not related to each other}. Besides, we explored also the possible influence of demographic factors and we did not find any result either.



\section{Threats to Validity}\label{sec:threats}

As any other empirical study, ours has faced several threats to validity that we report in this section.

\emph{Convenience Sampling.} Sampling is a major challenge in empirical research, especially when conducting surveys. Selecting a representative random sample of the target population is frequently beyond of what is possible in a research project. Therefore, researchers often opt for convenience sampling instead. This was also the case for our research. As we did not have the resources required to survey a large enough random sample of people, we selected our respondents by reaching out to a population related to our (social and industrial) networks.
We mitigated related threats in accordance with Gravetter and Forzano~\cite{gravetter2018research} by, first, carefully selecting a broad cross-section of the target population~\cite{baltes2016worse} and, second, being transparent in our demographics.

\emph{Quality of the instrument.} One of the main issues was to implement a balance on the amount of information gathered and the questionnaire completion time. As reported by Dillman et al.: ``The desire of surveyors to obtain answers to increasingly detailed questions needed for complex modeling of attitudes and behaviors appears often to be in conflict with the limitations and patience of respondents for providing answers” \cite{Dillman2014}. A further challenge was to ensure that the respondents all had the same understanding of the questions as we had when implementing them, also due to the verbose notion of what a standard eventually is. This threat to the internal validity was evident in the suggestion of standards that respondents gave in RQ1.1, which in general we discarded as the suggestions did not match the concept as we defined it.

Also in terms of construct validity, a related challenge also concerns the pre-selection of standards. To mitigate this threat, we conducted pilots of the questionnaire to ensure its sufficient completeness, but also the correct understanding and to find possible defects. Based on these pilots, we made some corrections and improvements (cf. Sect.~\ref{subsec:instrument}). An argument to support the adequacy of the final set of pre-selected standards is that respondents mentioned very few additional standards in the open text fields. However, there is always the possibility that some standard or type of standards is missing, for instance because respondents may not be aware that they actually use a standard or because they do not know the name even if they know they are using it.

\emph{Accuracy of the coding.} RQ3 and RQ4 were implemented with free-text questions. Therefore, we needed to apply content analysis. In order to reduce the threat of inaccurate coding, also in the light of coding being always a creative task influenced by researchers' bias, we applied some well-defined steps as reported in Sect.~\ref{sec:DataAn}. Designated authors took the lead of the process, iteratively identifying terms, topics, and categories.
The entire team discussed the results when a consolidated proposal was available. In this process, the codes were slightly adjusted to consolidate each other's opinions. Furthermore, in RQ4, we deemed convenient a last iteration involving a different author. As a final mitigation strategy to threats to the internal validity, we decided to disclose our raw data and the resulting coding scheme as a replication package \footnote{\url{http://doi.org/10.5281/zenodo.4755756}} so that other researchers can comprehend how we have drawn our conclusions from the data.

\emph{Average response rate.} We implemented RQ1 and RQ2 in the questionnaire through closed questions. Together with the dynamic implementation of the RQ2-related part of the questionnaire, almost all respondents provided answers to the questions related to RQ1 and RQ2. On the contrary, RQ3 and RQ4 were implemented in the questionnaire as open-ended free text questions, and this motivated that the percentage of responses dropped down to almost half of the survey population resulting in a threat to the internal validity, but also to the conclusions we can draw based on these two research questions.

\emph{Generalization of results.} Given the relatively low number of responses received and also the great diversity of respondents, we cannot claim that the findings of our study would necessarily correspond to results obtained when conducting the same survey with a different part of the RE population. Given the lack of evidence in the subject under investigation, we also cannot generalize our results by analogy.  This threatens the external validity of our results. However, we believe that the results already provide great insights into the current state of affairs in the adoption of RE standards and our disclosed replication package allows other researchers in the community to replicate our study with further samples.


\section{Conclusions and Future Work}\label{sec:conclusions}
In this article, we have reported on the results of a survey with practitioners on the use of RE-related standards in industry.  To our knowledge, this is the first study of this kind. 

We found that the knowledge and use of standards by the respondents of our study is much lower than anticipated. The majority of respondents decide themselves to use standards; mandatory use imposed by the company, a customer, or a regulator are minor factors in comparison. Beyond insufficient knowledge, there are also cultural and organizational factors that impede the widespread adoption of RE-related standards.

The interpretation of our findings is subject to future work. Our results could be an indicator that RE as a discipline is still in an immature state. However, the reason behind our observations could also be that RE is a discipline where standards are less important for systematic and successful work than they are in other engineering disciplines. 

Our findings are useful for a variety of stakeholders.
\begin{compactitem}

\item \emph{Practitioners} get an overview of RE-related standards and how their colleagues apply them in practice. This can alert practitioners to standards that are worth adopting. In addition, discussing major impediments and perceived advantages can help to plan for adoption. By promoting the systematic application of standards, our work as a whole is a means of improving the state of practice.

\item \emph{Standardization bodies} may learn about the acceptance of their standards by industry and understand what aspects could be improved.

\item \emph{Researchers} can take these results as a basis for further studies that will help interpret our findings and extend the research. Based on our disclosed open access replication package, they can even start their own analysis using our shared data set. It would also be interesting to repeat this study with a broader scope and a larger community of participants. Furthermore, it would be worthwhile to study factors that impede the adoption of standards in more detail and with different empirical methods, for example, by analyzing the influence of organizational culture. Finally, exploring the benefits associated with the use of standards could provide valuable insights for researchers and practitioners and promote informed decision-making about the adoption of standards. 

\end{compactitem}

The low degree of knowledge of relevant RE standards is an indicator that many requirements engineers feel that standards do not add sufficient value in the context of their work. In this light,
our results also provide a starting point for investigations about how to improve or augment existing standards so that they provide more value to the RE professionals who apply and use them.

\section*{Acknowledgements}
We thank our respondents for their time and effort which made this study possible. This work was partially supported by the KKS foundation through the S.E.R.T. Research Profile project at Blekinge Institute of Technology.

\bibliographystyle{plain}
\bibliography{re19standards}

\begin{IEEEbiography}[{\includegraphics[width=1in,height=1.25in,clip,keepaspectratio]{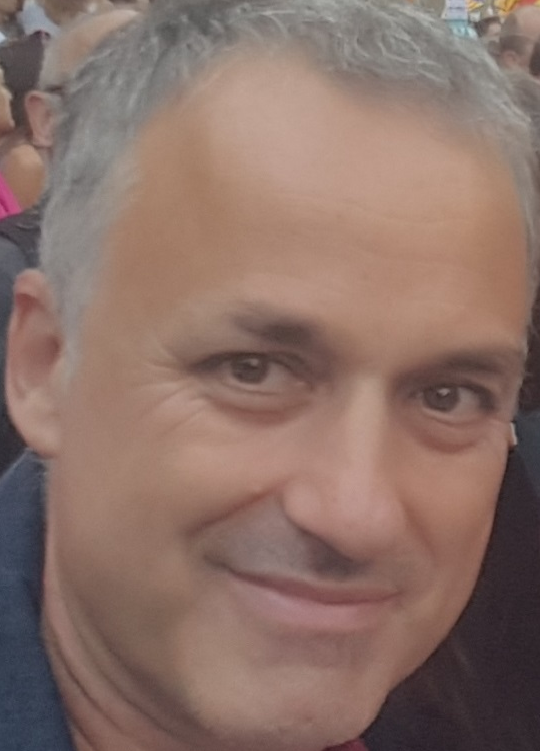}}]{Xavier Franch} received the PhD degree from the Universitat Polit\`ecnica de Catalunya, Spain. He is  a full professor with the Universitat Polit\`ecnica de Catalunya, Spain. His interests include requirements engineering, empirical software engineering and software quality. He is or has been member of IST, JSS, REJ, Computing and IET Software boards, and program co-chair of RE'16, ICSOC'14 and CAiSE'12. He is Full Member of IREB and member of ISERN. For more information, please visit \url{https://www.essi.upc.edu/~franch/}.
\end{IEEEbiography}

\begin{IEEEbiography}[{\includegraphics[width=1in,height=1.25in,clip,keepaspectratio]{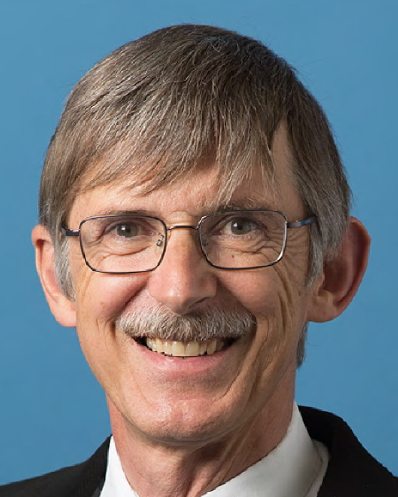}}]{Martin Glinz} received the Dr rer nat in computer science from RWTH Aachen University, Germany. He is a full professor emeritus at the University of Zurich, Switzerland. Before joining the University of Zurich, he worked in industry for 10 years. His interests include requirements and software engineering -- in particular modeling, validation, quality, and education. He is a member of the IEEE Computer Society and a full member of IREB, where he chairs the IREB Council. For more information, please visit \url{https://www.ifi.uzh.ch/~glinz}.
\end{IEEEbiography}

\begin{IEEEbiography}[{\includegraphics[width=1in,height=1.25in,clip,keepaspectratio]{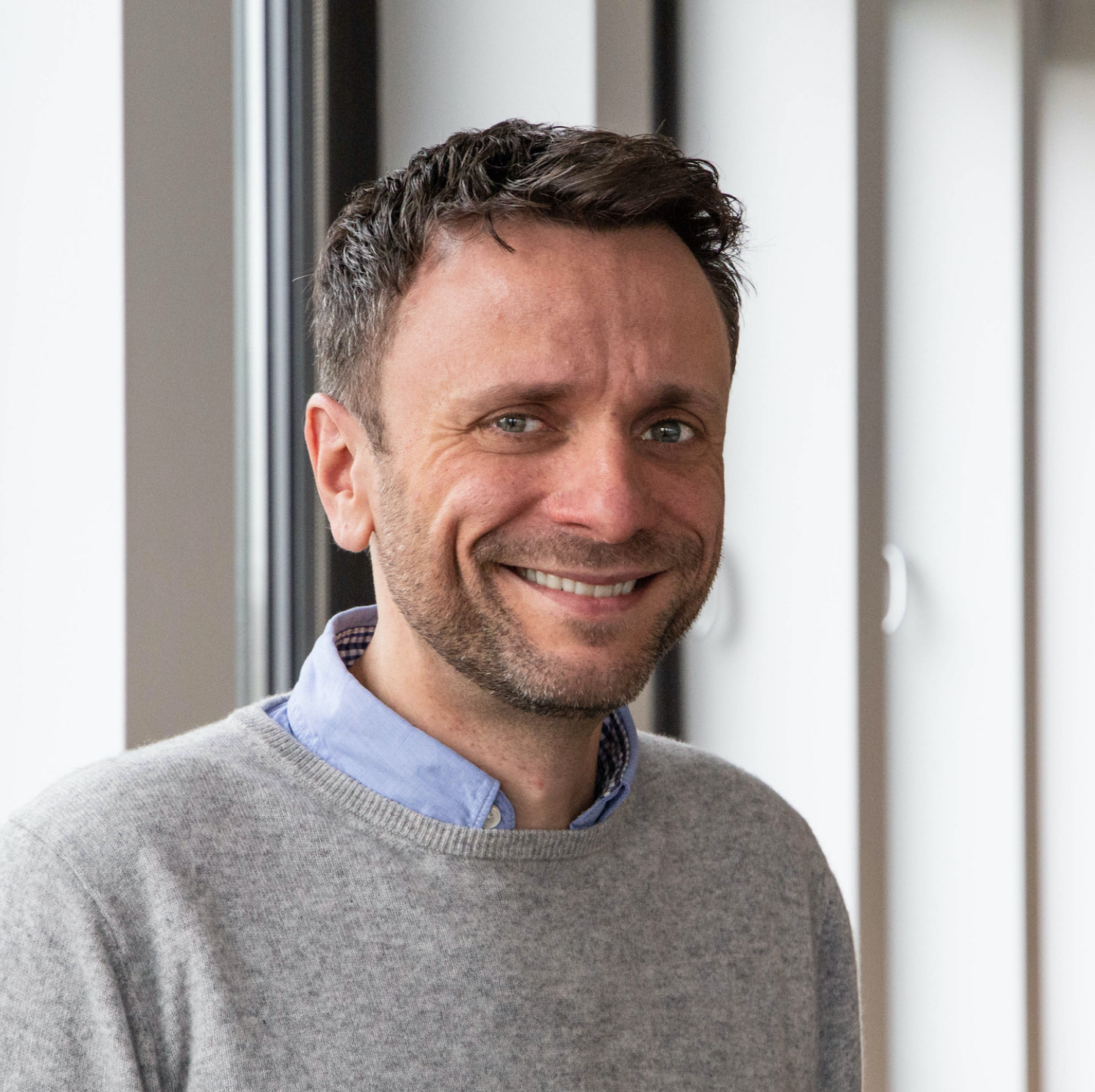}}]{Daniel Mendez} received the Dr rer nat and habilitation degree from the Technical University of Munich, Germany. He is full professor with the Blekinge Institute of Technology, Sweden, and Senior Scientist heading the research division Requirements Engineering at fortiss, Germany. He is further associate at the Technical University of Munich, Germany. His research concentrates on empirical software engineering with a particular focus on qualitative research in requirements engineering. For more information, please visit \url{http://www.mendezfe.org}.
\end{IEEEbiography}

\begin{IEEEbiography}[{\includegraphics[width=1in,height=1.25in,clip,keepaspectratio]{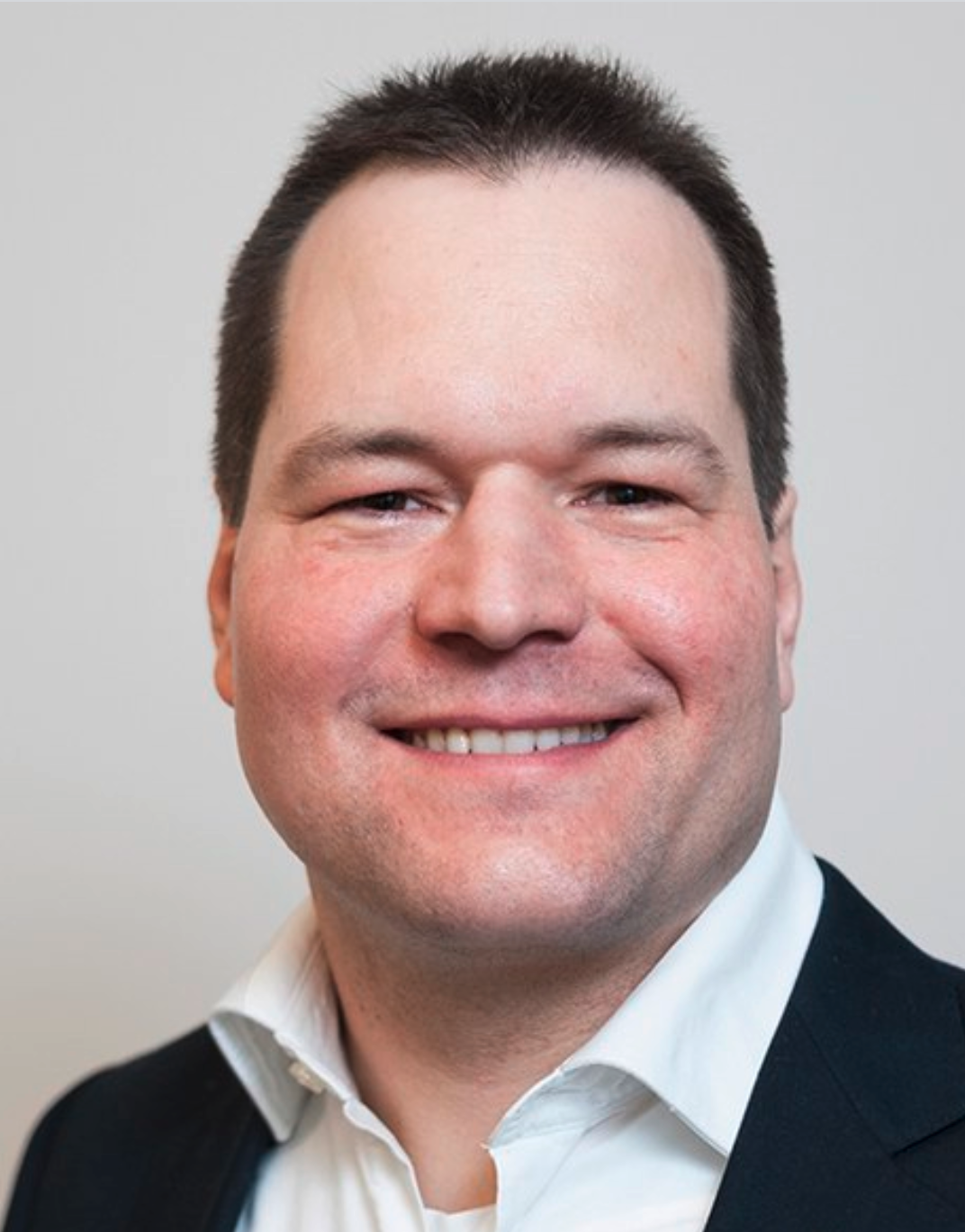}}]{Norbert Seyff} received the Dr techn in computer science from Johannes Kepler University Linz, Austria. He is professor with the University of Applied Sciences and Arts Northwestern Switzerland and senior research associate at the University of Zurich, Switzerland. His research interests include requirements engineering and sustainability design. He is a full member of IREB. For more information, please visit \url{https://www.fhnw.ch/en/people/norbert-seyff}.
\end{IEEEbiography}

\end{document}